%% file: text-main.tex
\documentclass[a4paper, fleqn]{cas-dc}
\usepackage[numbers]{natbib}
\usepackage[T1]{fontenc} % fix many problems regarding font encoding
\usepackage[utf8]{inputenc}
\usepackage[english]{babel} % proper hyphenation
\usepackage{tabularx}
\usepackage{booktabs}
\usepackage[flushleft]{threeparttable}
\usepackage{subcaption}
\usepackage{cuted,tcolorbox}
\usepackage{amsmath} % equations
\usepackage{listings} % equations
\usepackage{textcomp} % fix the \textquotesingle problem libs

\lstdefinestyle{code}{
backgroundcolor=\color{white},   % choose the background color; you must add \usepackage{color} or \usepackage{xcolor}; should come as last argument
basicstyle=\ttfamily,        % the size of the fonts that are used for the code
breakatwhitespace=false,         % sets if automatic breaks should only happen at whitespace
breaklines=true,                 % sets automatic line breaking
captionpos=b,                    % sets the caption-position to bottom
commentstyle=\color{gray},    % comment style
escapeinside={\%*}{*)},          % if you want to add LaTeX within your code
frame=lines,	                   % adds a frame around the code
keepspaces=true,                 % keeps spaces in text, useful for keeping indentation of code (possibly needs columns=flexible)
deletekeywords={time, anova},            % if you want to delete keywords from the given language
morekeywords={shapiro, lmer, Anova, wilcox, cliff},            % if you want to add more keywords to the set
numbers=none,                    % where to put the line-numbers; possible values are (none, left, right)
numbersep=5pt,                   % how far the line-numbers are from the code
numberstyle=\tiny\color{gray}, % the style that is used for the line-numbers
rulecolor=\color{black},         % if not set, the frame-color may be changed on line-breaks within not-black text (e.g. comments (green here))
showspaces=false,                % show spaces everywhere adding particular underscores; it overrides 'showstringspaces'
showstringspaces=false,          % underline spaces within strings only
showtabs=false,                  % show tabs within strings adding particular underscores
stepnumber=2,                    % the step between two line-numbers. If it's 1, each line will be numbered
tabsize=2,	                   % sets default tabsize to 2 spaces
title=\lstname,                   % show the filename of files included with \lstinputlisting; also try caption instead of title
}

\begin{document}
\newcommand{\Foutse}[1]{\textcolor{blue}{{\it [Foutse says: #1]}}}

\let\WriteBookmarks\relax
\def\floatpagepagefraction{1}
\def\textpagefraction{.001}
\shorttitle{A Large Scale Empirical Study of the Impact of Spaghetti Code and Blob Anti-patterns on Program Comprehension}
\shortauthors{Politowski et~al.}

\title[mode=title]{A Large Scale Empirical Study of the Impact of Spaghetti Code and Blob Anti-patterns on Program Comprehension}
\tnotemark[1]

\tnotetext[1]{Replication files: \url{https://doi.org/10.5281/zenodo.3601564}.}

\author[1]{Cristiano Politowski}[orcid=0000-0002-0206-1056] \cormark[1]
\ead{c_polito@encs.concordia.ca}

\author[2]{Foutse Khomh}[orcid=0000-0002-5704-4173]
\ead{foutse.khomh@polymtl.ca}

\author[3]{Simone Romano}[orcid=0000-0003-4880-3622]
\ead{simone.romano@uniba.it}

\author[4]{Giuseppe Scanniello}[orcid=0000-0003-0024-7508]
\ead{giuseppe.scanniello@unibas.it}

\author[5]{Fabio Petrillo}[orcid=0000-0002-8355-1494]
\ead{fabio@petrillo.com}

\author[1]{Yann-Ga\"el Gu\'{e}h\'{e}neuc}[orcid=0000-0002-4361-2563]
\ead{yann-gael.gueheneuc@encs.concordia.ca}

\author[6]{Abdou Maiga}
\ead{ma_karim@yahoo.fr}

\address[1]{Concordia University, Montreal, Quebec, Canada}
\address[2]{Polytechnique Montr\'{e}al, Montreal, Quebec, Canada}
\address[3]{University of Bari, Bari, Italy}
\address[4]{University of Basilicata, Potenza, Italy}
\address[5]{Université du Québec à Chicoutimi, Chicoutimi, Quebec, Canada}
\address[6]{Université Félix Houphouet Boigny, Abidjan, Ivory Coast}

\cortext[cor1]{Corresponding author}

\begin{abstract}
Context:
Several studies investigated the impact of anti-patterns (\textit{i.e.,} ``poor'' solutions to recurring design problems) during maintenance activities and reported that anti-patterns significantly affect the developers' effort required to edit files. However, before developers edit files, they must understand the source code of the systems. This source code must be easy to understand by developers.
\noindent \\
Objective:
In this work, we provide a complete assessment of the impact of two instances of two anti-patterns, Blob or Spaghetti Code, on program comprehension.
\noindent \\
Method:
We analyze the impact of these two anti-patterns through three empirical studies conducted at Polytechnique Montréal (Canada) with 24 participants; at Carlton University (Canada) with 30 participants; and at University Basilicata (Italy) with 79 participants.
\noindent \\
Results:
We collect data from 372 tasks obtained thanks to 133 different participants from the three universities. We use three metrics to assess the developers' comprehension of the source code: (1) the duration to complete each task; (2) their percentage of correct answers; and, (3) the NASA task load index for their effort.
\noindent \\
Conclusions:
We report that, although single occurrences of Blob or Spaghetti code anti-patterns have little effect on code comprehension, two occurrences of either Blob or Spaghetti Code significantly increases the developers' time spent in their tasks, reduce their percentage of correct answers, and increase their effort. Hence, we recommend that developers act on both anti-patterns, which should be refactored out of the source code whenever possible. We also recommend further studies on combinations of anti-patterns rather than on single anti-patterns one at a time.
\end{abstract}

\begin{keywords}
Anti-patterns \sep Blob \sep Spaghetti Code \sep Program Comprehension \sep Java
\end{keywords}

\maketitle

\input{sec-introduction}
\input{sec-rel-abbes}
\input{sec-method}
\input{sec-results}
\input{sec-discussion}
\input{sec-threats}
\input{sec-related-works}
\input{sec-conclusion}

\section*{Acknowledgements}
The authors thank all the anonymous participants for their time and efforts. The authors have been partly supported by the NSERC Discovery Grant program and Canada Research Chairs program as well as a FRQ-NT team project grant.

\balance
\bibliographystyle{cas-model2-names}
\bibliography{lib-main.bib,lib-review.bib}

%\vskip3pt

% \bio{pic-bio-cris}
% Author biography with author photo.
% Author biography. Author biography. Author biography.
% Author biography. Author biography. Author biography.
% Author biography. Author biography. Author biography.
% Author biography. Author biography. Author biography.
% Author biography. Author biography. Author biography.
% Author biography. Author biography. Author biography.
% Author biography. Author biography. Author biography.
% Author biography. Author biography. Author biography.
% Author biography. Author biography. Author biography.
% \endbio
%
%
% \bio{}
% Author biography without author photo.
% Author biography. Author biography. Author biography.
% Author biography. Author biography. Author biography.
% Author biography. Author biography. Author biography.
% Author biography. Author biography. Author biography.
% Author biography. Author biography. Author biography.
% Author biography. Author biography. Author biography.
% Author biography. Author biography. Author biography.
% Author biography. Author biography. Author biography.
% Author biography. Author biography. Author biography.
% \endbio
\end{document}

%% file: sec-introduction.tex
% !TeX root = text-main.tex
% !TeX spellcheck = en_US
% !TeX encoding = UTF-8%
\section{Introduction}
\label{sec:intro}

Program comprehension is central to an effective software maintenance and evolution \citep{Mayrhauser1995}. In his theory of program comprehension, Brooks~\cite{BROOKS1983} propound that developers understand software programs in a top-down manner; formulating hypotheses about the domain of the program, mapping this knowledge to the source code, and refining it incrementally. Therefore, understanding the factors affecting developers’s comprehension of the source code is essential to improve program comprehension and consequently software maintenance and evolution activities~\citep{Abbes2011}.

Anti-patterns, which are ``poor'' solutions to recurrent design problems, have been reported to significantly affect the effort required to explore and edit files \citep{Soh2016}. Furthermore, the increase in the number of anti-patterns in a system is likely to generate faults \citep{Ambros2010}. One may therefore question whether certain anti-patterns and--or combinations thereof significantly impede code understandability.
To understand whether and how anti-patterns affect code understandably, \citet{Abbes2011} conducted three experiments, with 24 participants each, investigating whether the occurrences of the
Blob\footnote{Blob anti-pattern is found in designs where one class monopolizes the processing, and other classes primarily encapsulate data. The major issue here is that the majority of the responsibilities are allocated to a single class \cite{Brown1998}, violating the single responsibility principle \cite{Martin2003}.}
and Spaghetti Code\footnote{Considering the Object Oriented paradigm, the software with Spaghetti Code anti-pattern may include a small number of objects that contain methods with very large implementations that invoke a single, multistage process flow \cite{Brown1998}.}
anti-patterns impact the understandability of source code for comprehension and maintenance tasks. They analyzed the Blob and Spaghetti Code anti-patterns individually and in combinations. They showed that the occurrence of one Blob or one Spaghetti Code in source code does not significantly reduce its understandability in comparison to source code without anti-pattern. However, they reported that the combination of one occurrence of the Blob anti-pattern with one of the Spaghetti Code significantly decreases understandability. They mentioned the following limitations to their study: (1) the experiments were conducted only with 24 students from the same computer-science program and (2) they did not test the impact of multiple occurrences of each anti-pattern individually. Thus, the results reported in this previous work are insightful but incomplete.

We extend this previous work \citep{Abbes2011} with two additional replications at two different universities. We conducted these replications with 30 students from Carleton University, Ottawa, Ontario, Canada, and 79 students from the University of Basilicata, Potenza, Italy, using six different Java software systems. In total, we collected data from 133 participants executing 372 tasks in three universities. We complemented the previous work with questions about the co-occurrences of the same anti-patterns: two occurrences of the Blob anti-patterns and two occurrences of the Spaghetti Code. We asked the participants to perform program comprehension tasks and we measured their performances using (1) the time that they spent performing their tasks; (2) their percentages of correct answers; and, (3) the NASA task load index of their effort.

Thus, we report in the following a complete, exhaustive study of the impact of the Blob and Spaghetti Code anti-patterns in isolation and in combinations. We show that the presence of two occurrences of either Blob or Spaghetti Code anti-pattern harms source-code understanding, weakening the performance of developers by increasing the time spent to finish the tasks, lowering the correctness of the answers, and increasing the effort to complete the tasks.

We bring further evidences with a larger experiment without setting aside previous results from \citet{Abbes2011}, which we describe in details in Section~\ref{sec:abbes}. Also, after presenting the new data in Section~\ref{sec:results}, we compare them with previous results in Section~\ref{sec:results-abbes}, in which Experiments \#1 (Blob) and \#2 (Spaghetti Code) correspond to the previous study. Thus, we name the new experiments Experiments \#3, \#4, \#5, and \#6.

The remainder of this paper is organized as follows: Section~\ref{sec:abbes} briefly describes the previous paper by \citet{Abbes2011}. Section~\ref{sec:method} describes the definition and design of our empirical studies. Section~\ref{sec:results} presents the study results from experiments with the Blob and Spaghetti Code anti-patterns as well the comparison with the previous data from \citet{Abbes2011}. Sections~\ref{sec:discussion} and~\ref{sec:threats} discuss the results and threats to their validity. Section~\ref{sec:relatedWork} relates our study with previous work. Section~\ref{sec:conclusion} concludes the paper with future work.

%% file: sec-rel-abbes.tex
% !TeX root = text-main.tex
% !TeX spellcheck = en_US
% !TeX encoding = UTF-8%
\section{Previous Work by \citet{Abbes2011}}
\label{sec:abbes}

\def\HAb{$H_{0_{1Blob}}$}
\def\HAs{$H_{0_{1SpaghettiCode}}$}
\def\HAm{$H_{0_{1Blob+1SpaghettiCode}}$}
\def\EAb{$E_{0_{1Blob}}$}
\def\EAs{$E_{0_{1SpaghettiCode}}$}
\def\EAm{$E_{0_{1Blob+1SpaghettiCode}}$}

\citet{Abbes2011} designed and conducted three experiments, with 24 participants each, to collect data on the performance of developers on tasks related to program comprehension and assessed the impact of anti-patterns: Blob and Spaghetti Code. The first and second experiments examined the impact of an occurrence of Blob and Spaghetti Code anti-pattern, independently, on system understandability. In the third experiment, they used a combination of one Blob and one Spaghetti Code anti-pattern to analyze its impact.

\paragraph{Research Question:} \citet{Abbes2011} addressed the following research questions: what is the impact of an occurrence of the Blob anti-pattern (respectively of the Spaghetti Code anti-pattern and of the two anti-patterns mixed) on understandability?

\paragraph{Objects of the Study:} Each experiment was performed using three software systems written in Java. \autoref{tab:exp1-2-sys} presents a short description of the systems.
For the sake of clarity, we are omitting \citet{Abbes2011} experiment \#3 here because they combined \emph{Blob+Spaghetti Code}, which we will not perform in our experiments (see next Section).

% \begin{table}[width=1\linewidth, pos=ht]
% \centering
% \caption{Object systems used in \citet{Abbes2011}.}
% \label{tab:exp1-2-sys}
% \begin{tabular}{@{}lrrr@{}}
% \toprule
% System & Ver. & NoC & SLOCs \\ \midrule
% YAMM & 0.9.1 & 64 & 11,272  \\
% JVerFileSystem & - & 167 & 38,480  \\
% AURA & - & 95 & 10,629  \\
% GanttProject & 2.0.6 & 527 & 68,545  \\
% JFreeChart & 1.0.13 & 989 & 302,844 \\
% Xerces & 2.7.0 & 740 & 233,331  \\ \bottomrule
% \end{tabular}
% \end{table}

\begin{table}[width=1\linewidth, pos=ht]
% \footnotesize
\caption{Object systems used in \citet{Abbes2011}.}
\label{tab:exp1-2-sys}
\begin{tabular*}{\tblwidth}{@{}lllrrr@{}}
\toprule
Exp & AP & System & Ver. & NoC & SLOC \\ \midrule
\#1 & Blob & YAMM & 0.9.1 & 64 & 11,272  \\
\#1 & Blob & JVerFileSystem & - & 167 & 38,480  \\
\#1 & Blob & AURA & - & 95 & 10,629  \\
\#2 & Spaghetti & GanttProject & 2.0.6 & 527 & 68,545  \\
\#2 & Spaghetti & JFreeChart & 1.0 & 989 & 302,844 \\
\#2 & Spaghetti & Xerces & 2.7.0 & 740 & 233,331  \\ \bottomrule
\end{tabular*}
\end{table}

From each system, they selected randomly a subset of classes responsible for a specific feature to limit the size of the displayed source code. They performed manual refactorings \citep{Fowler2012} on each subset of each system to remove all other occurrences of (other) anti-patterns to reduce possible bias by other anti-patterns, while keeping the systems compiling and functional.

Finally, they refactored each subset of each system to obtain a subset of classes in which there was no occurrence of any anti-pattern. This subset was used as a base line to compare the participants’ performance.

\paragraph{Variables of the Study:} The independent variables for the first, second, and third experiments are an occurrence of Blob, an occurrence of Spaghetti Code, and a co-occurrence of Blob and Spaghetti Code, respectively.
The dependent variables of the three experiments are: participants' performance, in terms of effort, time spent, and percentage of correct answers. The effort of a participant was measured using the NASA Task Load Index (TLX) \citep{Hart1988}. They also investigated mitigating variables: participants knowledge level in Java; participant's knowledge level of Eclipse; and, participant's knowledge level in software engineering. They assessed the participants' knowledge levels using a post-mortem questionnaire that was administered to every participant at the end of the experiments.

\paragraph{Participants:} Each experiment was performed by 24 anonymous participants. Some participants were enrolled in the M.Sc.\ and Ph.D.\ programs in computer and software engineering at Polytechnique Montréal or in computer science at Université de Montréal. Others were professionals working for software companies in the Montréal area, recruited through the authors' industrial contacts.

\paragraph{Experiment Questions:} The questions used to elicit comprehension tasks and collect data on the participants' performances were related to: (1) finding a focus point in some subset of the classes and interfaces of some source code, relevant to a comprehension task; (2) focusing on a particular class believed to be related to some task and on directly-related classes; and, (3) understanding a number of classes and their relations in some subset of the source code.

\paragraph{Experiment Design:}
The design for each experiment is presented in \autoref{tab:design-abbes}.
An Eclipse workspace was provided to each participant. This workspace contained a compiling and functional subset of classes, linked to JAR files containing the rest of the compiled code. It also contained a timer, a TLX program, a brief tutorial on Eclipse, a brief description of the system at hand, and the post-mortem questionnaire used to measure the participant knowledge. The three experiments were conducted in the same laboratory, with the same computer and software environments. The participants did not have prior knowledge of the systems on which they performed the comprehension tasks.

\begin{table}[width=1\linewidth, pos=ht]
\caption{Design of Experiments \#1 and \#2 \citep{Abbes2011}, with Blob and Spaghetti Code anti-pattern, showing the participants IDs.}
\label{tab:design-abbes}
\begin{tabular*}{\tblwidth}{@{}lLrrrrrrrr@{}}
\toprule
Exp & AP & \multicolumn{8}{l}{Participants' ID} \\ \midrule
\#1 & Blob &  1 & 2 & 6 & 14 & 15 & 17 & 20 & 22 \\
\#1 & Blob &  3 & 7 & 9 & 11 & 12 & 18 & 21 & 24 \\
\#1 & Blob &  4 & 5 & 8 & 10 & 13 & 16 & 19 & 23 \\
\#1 & - &  4 & 7 & 9 & 11 & 13 & 18 & 19 & 23 \\
\#1 & - &  1 & 5 & 8 & 10 & 15 & 16 & 20 & 22 \\
\#1 & - &  2 & 3 & 6 & 12 & 14 & 17 & 21 & 24 \\ \addlinespace
\#2 & Spaghetti & 1 & 2 & 6 & 14 & 15 & 17 & 20 & 22 \\
\#2 & Spaghetti & 3 & 7 & 9 & 11 & 12 & 18 & 21 & 24 \\
\#2 & Spaghetti & 4 & 5 & 8 & 10 & 13 & 16 & 19 & 23 \\
\#2 & - & 4 & 7 & 9 & 11 & 13 & 18 & 19 & 23 \\
\#2 & - & 1 & 5 & 8 & 10 & 15 & 16 & 20 & 22 \\
\#2 & - & 2 & 3 & 6 & 12 & 14 & 17 & 21 & 24 \\ \bottomrule
\end{tabular*}
\end{table}

\begin{tcolorbox}[title=Summary: Results of \citet{Abbes2011}]
Results of the three experiments showed that the occurrence of one Blob or one Spaghetti Code anti-pattern in the source code of a software system \emph{does not} significantly make its comprehension harder for participants when compared to source code without anti-pattern. However, the combination of one Blob and one Spaghetti Code anti-pattern impacts negatively and significantly a systems' understandability.
\end{tcolorbox}

%% file: sec-method.tex
% !TeX root = text-main.tex
% !TeX spellcheck = en_US
% !TeX encoding = UTF-8%
\section{Experiment Design}
\label{sec:method}

The previous section presented the results of two experiments, considered here \#1 and \#2. This paper aims to extend that work by adding four more experiments, called \#3, \#4, \#5, and \#6, in two different locations.
Each one of these deal with two occurrences of Blob and two occurrences of Spaghetti Code in a software system.
Our experiment does not consider combinations of anti-patterns, for example \emph{Blob + Spaghetti Code}, as \citet{Abbes2011} did.

\autoref{tab:exps} shows the experiments and its participants. In each experiment, we assign two systems to each participant: one containing two occurrences of the Blob or Spaghetti Code anti-pattern and one without any anti-pattern. We then measure and compare the participants' performances for both systems in term of program comprehension. Thus, we can determine the impact of the two anti-patterns on understandability, from the point of view of developers, in the context of Java systems.

\begin{table*}[width=1\textwidth,cols=4,pos=ht]
% \footnotesize
\begin{threeparttable}
\caption{Group of experiments.}
\label{tab:exps}
\centering
\begin{tabular*}{\tblwidth}{@{}Llcccl@{}}
\toprule
Location & AP Type & \#APs & Experiment ID & Participants & Data Origin \\ \midrule
Montréal, Canada & Blob & 1 & \#1 & 24 & \citet{Abbes2011} \\
Montréal, Canada & Spaghetti Code & 1 & \#2 & 24 & \citet{Abbes2011} \\  \addlinespace
Ottawa, Canada & Blob & 2 & \#3 & 30 & This paper \\
Ottawa, Canada & Spaghetti Code & 2 & \#4 & 29 & This paper \\ \addlinespace
Potenza, Italy & Blob & 2 & \#5 & 41 & This paper \\
Potenza, Italy & Spaghetti Code & 2 & \#6 & 38 & This paper \\ \bottomrule
\end{tabular*}
\begin{tablenotes}
\small
\item \#APs means how many occurrences of the anti-pattern are in the source code.
\end{tablenotes}
\end{threeparttable}
\end{table*}

\subsection{Research Question}

Our research question stems from our goal of understanding the impact of two occurrences of Blob and Spaghetti Code anti-patterns on program comprehension.
The research questions that our study addresses are:

\begin{itemize}
\item \textbf{RQ1}: What is the impact of two occurrences of the Blob anti-pattern on the participants' \emph{average time spent} understanding a software system?
\item \textbf{RQ2}: What is the impact of two occurrences of the Blob anti-pattern on the participants' \emph{average correct answers} to understanding questions?
\item \textbf{RQ3}: What is the impact of two occurrences of the Blob anti-pattern on the participants' \emph{average effort spent} understanding a software systems?
\item \textbf{RQ4}: What is the impact of two occurrences of the Spaghetti Code anti-pattern on the participants' \emph{average time spent} understanding software systems?
\item \textbf{RQ5}: What is the impact of two occurrences of the Spaghetti Code anti-pattern on the participants' \emph{average correct answers} to understanding questions?
\item \textbf{RQ6}: What is the impact of two occurrences of the Spaghetti Code anti-pattern on the participants' \emph{average effort spent} understanding a software systems?
\end{itemize}

\subsection{Hypotheses}
\label{sec:hyp}

From Research Questions 1, 2, and 3, we formulate the following \textit{null} hypotheses for the Blob anti-pattern:

\begin{itemize}
\item $H_{{BlobTime}}$ There is \emph{no} statistically significant difference between the participants' \emph{average time spent} when executing comprehension tasks on source code containing two occurrences of the \emph{Blob anti-pattern} compared to participants executing the same tasks on source code without any occurrences of the Blob anti-pattern.

\item $H_{{BlobAnswer}}$ There is \emph{no} statistically significant difference between the participants' \emph{correctness of the answers} when executing comprehension tasks on source code containing two occurrences of the \emph{Blob anti-pattern} compared to participants executing the same tasks on source code without any occurrences of the Blob anti-pattern.

\item $H_{{BlobEffort}}$ There is \emph{no} statistically significant difference between the participants' \emph{effort} when executing comprehension tasks on source code containing two occurrences of the \emph{Blob anti-pattern} compared to participants executing the same tasks on source code without any occurrences of the Blob anti-pattern.
\end{itemize}

For the Spaghetti Code anti-pattern and on the basis of Research Questions 4, 5 and 6, we formulate the following \textit{null} hypotheses:

\begin{itemize}
\item $H_{{SpaghettiTime}}$ There is \emph{no} statistically significant difference between the participants' \emph{average time spent} when executing comprehension tasks on source code containing two occurrences of the \emph{Spaghetti Code anti-pattern} compared to participants executing the same tasks on source code without any occurrences of the Spaghetti Code anti-pattern.

\item $H_{{SpaghettiAnswer}}$ There is \emph{no} statistically significant difference between the participants' \emph{correctness of the answers} when executing comprehension tasks on source code containing two occurrences of the \emph{Spaghetti Code anti-pattern} compared to participants executing the same tasks on source code without any occurrences of the Spaghetti Code anti-pattern.

\item $H_{{SpaghettiEffort}}$ There is \emph{no} statistically significant difference between the participants' \emph{effort} when executing comprehension tasks on source code containing two occurrences of the \emph{Spaghetti Code anti-pattern} compared to participants executing the same tasks on source code without any occurrences of the Spaghetti Code anti-pattern.
\end{itemize}

\subsection{Objects}

We choose three systems for each experiment, all developed in Java, and briefly described in Table~\ref{tab:systems}. We performed each experiment on the three different systems, because one system could be intrinsically either easier or harder to understand.

\begin{table*}[width=1\textwidth,cols=4,pos=ht]
\begin{threeparttable}
\caption{Object Systems.}
\label{tab:systems}
\begin{tabular*}{\tblwidth}{@{}Lllrrrrr@{}}
\toprule
Experiment & Anti-pattern & System & Version & NoC & SLOCs & Release & Commits \\ \midrule
\#3 and \#5 & Blob & Azureus & 2.3.0.6 & 1,449 & 191,963 & 2005 & 27300 \\
\#3 and \#5 & Blob & iTrust & 11 & 565 & 21,901 & 2010 & 256 \\
\#3 and \#5 & Blob & SIPComm & 1 & 1,771 & 486,966 & 2010 & 12693 \\ \addlinespace
\#4 and \#6 & Spaghetti Code & ArgoUml & 0.2 & 1,230 & 113,017 & 2006 & 16144 \\
\#4 and \#6 & Spaghetti Code & JHotDraw & 5.4b2 & 484 & 72,312 & 2004 & 765 \\
\#4 and \#6 & Spaghetti Code & Rhino & 1.6R5 & 108 & 48,824 & 2009 & 3532 \\ \bottomrule
\end{tabular*}
\begin{tablenotes}
\small
\item \#Commits is the approximated number of commits of each repository, gathered on December 13$^th$, 2018. Some of the repositories where migrated, for example from SVN to Github and some historical data may have been lost.
\end{tablenotes}
\end{threeparttable}
\end{table*}

For Experiment \#3 and \#5, we use Azureus\footnote{\url{http://www.vuze.com/}}, a Bit Torrent client used to transfer files via the Bit Torrent protocol (now known as Vuze); iTrust\footnote{\url{http://agile.csc.ncsu.edu/iTrust/}}, a medical application that provides patients with a means to keep up with their medical history and records as well as communicate with their doctors; and, SIP\footnote{\url{http://www.jitsi.org/}}, an audio/video Internet phone and instant messenger that supports some of the most popular instant messaging and telephony protocols (now known as Jitsi).

For Experiment \#4 and \#6, we use ArgoUML\footnote{\url{http://argouml.tigris.org/}}, an UML diagramming application written in Java; JHotDraw\footnote{\url{http://www.jhotdraw.org/}}, a graphic framework for drawing 2D graphics; and, Rhino\footnote{\url{http://www.mozilla.org/rhino/}}, an open-source JavaScript interpreter.

We used the following criteria to select the systems. First, we selected open-source systems; therefore other researchers can replicate our experiment. Second, we avoided to select small systems that do not represent the ones on which developers work normally. We also chose these systems because they are typical examples of systems having continuously evolved over periods of time of different lengths. Hence, the occurrences of Blob and Spaghetti Code in these systems are not coincidence but are realistic. 
We use the anti-pattern detection technique DETEX, which stems from the DECOR method \citep{Moha2005,Moha2010} to ensure that each system has at least two occurrences of the Blob and--or the Spaghetti Code anti-pattern. We randomly assigned a set of three systems to each experiment.

We manually validated the detected occurrences. From each system, we selected randomly a subset of classes responsible for managing a specific feature to limit the size of the source code given to the participants. For example, in iTrust, we chose the source code of the classes providing patients with access to their medical history and records.

The difference between systems would not impact our results because, regardless of the sizes, a participant concentrates her efforts only on a small part of the subset of the source code in which the anti-pattern class plays a \emph{central} role, \textit{i.e.}, the Blob class and its surrounding classes. Therefore, we ensure that all participants perform comprehension tasks within, almost, the same piece of code. Then, we refactor each subset of each system to remove all other occurrences of (other) anti-patterns to reduce possible bias by other anti-patterns, while keeping the system compiling and functioning. We perform manual refactoring following the guidelines from \citet{Fowler2012}. For example, when dealing with a Blob class, we replace it by multiple smaller classes and move some methods to existing/new classes.

For experiments \#3 and \#5, each subset contains two occurrences of the Blob anti-pattern. For Experiment \#4 and \#6, each subset contains two occurrences of the Spaghetti Code anti-pattern. We refactor each subset of the systems to obtain new subsets in which no occurrence of the anti-patterns exist. We use these subsets as baselines to compare the participants' performances and test our \textit{null} hypotheses.

\subsection{Independent and Dependent Variables}

The \emph{independent} variables are those variables that we can control. The variables should have some effect on the dependent variables and must be controllable \citep{Wohlin2012}. These are the independent variables of our experiments:

\begin{itemize}
\item \emph{Experiment ID} is a nominal variable identifying the experiment. For example, a value for this variable equal to \textit{\#3} identifies the third experiment.

\item \emph{Object} is a nominal variable representing the system. For example, it assumes the values \textit{Azureus}, \textit{iTrust}, and so on.

\item \emph{Treatment} is a nominal variable indicating whether or not the source code contains an anti-pattern. For example, in the experiments \#3 and \#5, this variable assumes as values \textit{Blob} or \textit{NoBlob}, \textit{i.e.,} source code with/out the Blob anti-pattern.
\end{itemize}

The \emph{dependent} variables measure the participants' performance, in terms of time spent, percentage of correct answers, and effort.

We measure the \emph{time} using a timer developed in Java that the participants must start before performing their comprehension tasks to answer the questions and stop when done. The time variable assumes values between 0 seconds and ``N'' seconds, where a value close to zero indicates that the task was performed quickly by the participant.

We compute the percentage of \emph{correct answers} for each question by dividing the number of correct elements found by a participant by the total number of correct elements they should have found. For example, for a question on the references to a given object, if there are ten references but the participant finds only four, the percentage would be forty. Then, we average these percentages to get an overall percentage for each participant. This variable assumes values between 0\% and 100\% where a value close to 100\% indicates a high correctness in the answers.

We measure the participants \emph{effort} using the NASA Task Load Index (TLX) \citep{Hart1988}. The TLX assesses the subjective workload of participants. It is a multi-dimensional measure that provides an overall workload index based on a weighted average of ratings on six sub-scales: mental demands, physical demands, temporal demands, own performance, effort, and frustration. NASA provides a computer program to collect weights of six sub-scales and ratings on these six sub-scales. We combine weights and ratings provided by the participants into an overall weighted workload index by multiplying ratings and weights; the sum of the weighted ratings divided by fifteen (sum of the weights) represents the effort \citep{TLX1988} with values between 0\% and 100\% where a value close to 100\% indicates a high effort for a participant.

Finally, \autoref{tab:data-sample} shows a sample of the stored data with the dependent and independent variables.

\begin{table*}[width=1\textwidth,cols=4,pos=ht]
\caption{Example of the data, the variables and its structure.}
\label{tab:data-sample}
\begin{tabular*}{\tblwidth}{@{}llllllLrrr@{}}
\toprule
N & Place & Exp & Global ID & Local ID & Object & Treatment & Time & Answer & Effort \\ \midrule
97 & Carleton & e3 & s25 & s1 & iTrust & Blob & 305 & 23 & 71 \\
98 & Carleton & e3 & s25 & s1 & Azureus & NoBlob & 85.7 & 67 & 57 \\
99 & Carleton & e3 & s26 & s2 & iTrust & Blob & 355 & 33 & 58 \\
100 & Carleton & e3 & s26 & s2 & Azureus & NoBlob & 143 & 72 & 48 \\
101 & Carleton & e3 & s27 & s3 & SIPComm & Blob & 273 & 33 & 57 \\
102 & Carleton & e3 & s27 & s3 & Azureus & NoBlob & 150 & 83 & 43 \\
103 & Carleton & e3 & s28 & s4 & SIPComm & Blob & 392 & 11 & 67 \\
104 & Carleton & e3 & s28 & s4 & iTrust & NoBlob & 95 & 78 & 52 \\
105 & Carleton & e3 & s29 & s5 & Azureus & Blob & 277 & 22 & 54 \\
106 & Carleton & e3 & s29 & s5 & iTrust & NoBlob & 92 & 89 & 49 \\ \bottomrule
\end{tabular*}
\end{table*}

\subsubsection{Participants' Technical Expertise}

The participants' knowledge levels were assessed using a post-mortem questionnaire that was administered to every participant at the end of its participation to an experiment. The questions were in five-point-scale, as described below. \autoref{tab:data-skills-sample} shows a subset of the data regarding the skills.

\begin{enumerate}
\item How do you rate your skills and knowledge in \emph{Software Engineering}?
\item How do you rate your level in \emph{Java}?
\item How do you rate your level in \emph{Eclipse}?
\end{enumerate}

\begin{table}[pos=ht]
\caption{Example of the data regarding the skills of the participants.}
\label{tab:data-skills-sample}
\begin{tabular*}{\tblwidth}{@{}lllllrrr@{}}
\toprule
Place & Exp & G.ID & L.ID & Treat. & SE & Java & Eclipse \\ \midrule
Montr. & e1 & s1 & s1 & Blob & 3 & 3 & 3 \\
Montr. & e1 & s2 & s2 & Blob & 3 & 3 & 3 \\
Montr. & e1 & s3 & s3 & Blob & 3 & 3 & 3 \\
Montr. & e1 & s4 & s4 & Blob & 3 & 3 & 3 \\
Montr. & e1 & s5 & s5 & Blob & 3 & 3 & 4 \\
Montr. & e1 & s6 & s6 & Blob & 3 & 5 & 3 \\ \bottomrule
\end{tabular*}
\end{table}

\subsection{Participants}

For each group of experiments shown in \autoref{tab:exps}, we use a random sampling to select and assign participants. The population is a convenient sampling of students participating in classes at two universities.

\paragraph{Experiments \#3 and \#4:} Each experiment was performed by 30 (30 for Blob tasks and 29 for Spaghetti Code tasks) anonymous participants, with some from the M.Sc.\ and Ph.D.\ programs of computer and software engineering at the computer science department of Carleton University.

\paragraph{Experiments \#5 and \#6:} 79 anonymous participants performed the experiments at the University of Basilicata, 41 out of 79 dealt with the Blob anti-pattern in the experiment \#5 and 38 dealt with the Spaghetti Code anti-pattern in the experiment \#6. Participants were enrolled to either the B.Sc.\ or the M.Sc.\ program at University of Basilicata. Seven participants were enrolled to the M.Sc.\ in the first year in computer engineering. The remaining students were enrolled in the third year in computer science. Different participants participated in each experiment.

\subsection{Questions}
\label{sec:questions}

We use comprehension questions to elicit comprehension tasks and collect data on the participants' performances. We consider questions in three of the four categories of questions regularly asked and answered by developers~\cite{Sillito2008}:

\begin{enumerate}
\item Finding a focus point in some subset of the classes and interfaces of some source code, relevant to a comprehension task;

\item Focusing on a particular class believed to be related to some task and on directly-related classes;

\item Understanding a number of classes and their relations in some subset of the source code;

\item Understanding the relations between different subsets of the source code. Each category contains several questions of the same type.
\end{enumerate}

We choose questions only in the first three categories, because the last category pertains to different subsets of the source code and, in our experiments, we focus only on one subset containing the occurrence(s) of the anti-pattern(s).

The six questions are the followings. The text in bold is a placeholder that we replace by appropriate behaviors, concepts, UI elements, methods, and types depending on the systems on which the participants perform their tasks.

\begin{itemize}
\item Category 1: Finding focus points:
\begin{itemize}
\item Question 1: Where is the code involved in the implementation of \textbf{[this behavior]}?

\item Question 2: Which type represents \textbf{[this domain concept]} or \textbf{[this UI element or action]}?
\end{itemize}

\item Category 2: Expanding focus points:
\begin{itemize}
\item Question 1: Where is \textbf{[this method]} called or \textbf{[this type]} referenced?

\item Question 2: What data can we access from \textbf{[this object]}?
\end{itemize}

\item Category 3: Understanding a subset:
\begin{itemize}
\item Question 1: How are \textbf{[these types or objects]} related?

\item Question 2: What is the behavior that \textbf{[these types]} provide together and how is it distributed over \textbf{[these types]}?
\end{itemize}
\end{itemize}

For example, with Azureus, we replace ``\textbf{this behavior}'' in Question 1, Category 1, by ``the health status of the resource to be downloaded, by calculating the number of seeds and peers'' and the question reads: ``Where is, in this project, the method involved in the implementation of the health status of the resource to be downloaded, by calculating the number of seeds and peers?''

For Category 2, it could seem that participants could simply answer the questions using Eclipse but they must identify and understand the classes or methods that they believe to be related to the task. Moreover, discovering classes and relationships that capture incoming connections prepare the participants for the questions of the third category.

Below are some examples of related questions for the ArgoUML system:

\begin{itemize}
\item \textit{Where is, in this project, the method involved in the implementation of parsing a line of text (containing the notation UML for an operation) to align the Operation to the specification given?}

\item \textit{What data can we access from an object of the type OperationNotationUml?}

\item \textit{How are the types OperationNotationUml and NotationUtilityUml related?}
\end{itemize}

\subsection{Design}

The distribution of the participants was random, as long as the same participant did not perform a task on the same system more than once to avoid learning biases. We have three different systems, each with two possibilities: containing or not containing the occurrences of an anti-pattern. Hence, six combinations are possible for each experiment. In each combination, we prepare a set of comprehension questions, which together form a \emph{treatment}.

\autoref{tab:design} shows the design of the experiments \#3, \#4, \#5 and \#6 for the Blob and Spaghetti Code anti-patterns, respectively, as well as the participants who performed the tasks and the respective systems.

\begin{table*}[width=1\textwidth, cols=4, pos=ht]
\caption{Design of the experiments \#3, \#4, \#5 and \#6, for the Blob and Spaghetti Code anti-patterns, respectively, showing the participants IDs.}
\label{tab:design}
\setlength{\tabcolsep}{3.6pt}
\begin{tabular*}{\tblwidth}{@{}llLrrrrrrrrrrrrrrr@{}}
\toprule
Exp & Anti-pattern & Object & \multicolumn{15}{l}{Participants' ID} \\ \midrule
\#3 & Blob & iTrust & 1 & 12 & 44 & 45 & 46 & 11 & 13 & 15 & 17 & 24 &  &  &  &  &  \\
\#3 & Blob & Azureus & 42 & 43 & 2 & 6 & 8 & 9 & 14 & 18 & 21 & 22 &  &  &  &  &  \\
\#3 & Blob & SIPComm & 23 & 33 & 3 & 4 & 5 & 7 & 10 & 16 & 19 & 20 &  &  &  &  &  \\ \addlinespace

\#3 & - & iTrust & 33 & 42 & 2 & 3 & 5 & 7 & 8 & 18 & 19 & 21 &  &  &  &  &  \\
\#3 & - & Azureus & 1 & 12 & 23 & 44 & 45 & 4 & 10 & 11 & 16 & 20 & 24 &  &  &  &  \\
\#3 & - & SIPComm & 43 & 46 & 6 & 9 & 13 & 14 & 15 & 17 & 22 &  &  &  &  &  &  \\ \addlinespace

\#5 & Blob & iTrust & 1 & 12 & 43 & 6 & 7 & 9 & 15 & 21 & 27 & 29 & 36 & 37 &  &  &  \\
\#5 & Blob & Azureus & 44 & 46 & 3 & 4 & 10 & 14 & 17 & 18 & 22 & 26 & 28 & 34 & 38 & 40 &  \\
\#5 & Blob & SIPComm & 33 & 42 & 45 & 2 & 8 & 11 & 16 & 19 & 20 & 24 & 30 & 31 & 35 & 39 & 41 \\ \addlinespace

\#5 & - & iTrust & 33 & 44 & 46 & 3 & 10 & 11 & 16 & 18 & 19 & 24 & 31 & 35 & 38 & 40 &  \\
\#5 & - & Azureus & 1 & 42 & 45 & 2 & 7 & 8 & 15 & 20 & 27 & 29 & 30 & 36 & 39 & 41 &  \\
\#5 & - & SIPComm & 12 & 43 & 4 & 6 & 9 & 14 & 17 & 21 & 22 & 26 & 28 & 34 & 37 &  &  \\ \midrule

\#4 & Spaghetti C. & ArgoUML & 1 & 33 & 45 & 2 & 6 & 7 & 10 & 11 & 15 & 19 &  &  &  &  &  \\
\#4 & Spaghetti C. & JHotDraw & 12 & 23 & 46 & 3 & 9 & 13 & 16 & 17 & 18 & 22 &  &  &  &  &  \\
\#4 & Spaghetti C. & Rhino & 42 & 43 & 44 & 4 & 5 & 8 & 14 & 20 & 21 &  &  &  &  &  &  \\ \addlinespace

\#4 & - & ArgoUML & 12 & 23 & 43 & 3 & 5 & 8 & 13 & 14 & 16 & 18 &  &  &  &  &  \\
\#4 & - & JHotDraw & 1 & 33 & 42 & 44 & 2 & 4 & 10 & 15 & 19 & 20 & 21 &  &  &  &  \\
\#4 & - & Rhino & 45 & 46 & 6 & 7 & 9 & 11 & 17 & 22 &  &  &  &  &  &  &  \\ \addlinespace

\#6 & Spaghetti C. & ArgoUML & 23 & 46 & 3 & 4 & 10 & 14 & 17 & 18 & 26 & 28 & 32 & 34 & 38 & 40 &  \\
\#6 & Spaghetti C. & JHotDraw & 1 & 43 & 6 & 9 & 13 & 15 & 21 & 25 & 27 & 29 & 36 &  &  &  &  \\
\#6 & Spaghetti C. & Rhino & 33 & 42 & 45 & 2 & 5 & 8 & 11 & 16 & 19 & 24 & 31 & 35 & 39 &  &  \\ \addlinespace

\#6 & - & ArgoUML & 1 & 42 & 45 & 2 & 8 & 13 & 15 & 25 & 27 & 29 & 36 & 39 &  &  &  \\
\#6 & - & JHotDraw & 33 & 46 & 3 & 5 & 10 & 11 & 16 & 18 & 19 & 24 & 31 & 32 & 35 & 38 & 40 \\
\#6 & - & Rhino & 23 & 43 & 4 & 6 & 9 & 14 & 17 & 21 & 26 & 28 & 34 &  &  &  &  \\ \bottomrule
\end{tabular*}
\end{table*}

\subsection{Procedure}

We received the agreement from the Ethical Review Boards of Université de Montréal and Carleton University to perform and publish this study. The collected data is anonymous. The participants could leave any experiment at any time, for any reason, and without penalty of any kind. No participant left the study or took more than 45 minutes to perform the experiment. The participants knew that they would perform comprehension tasks but did know the goal of the experiment or the system that they were studying or whether the system contained or not anti-pattern(s). We informed participants of the goal of the study after collecting their data, before they left their experiments.

For each experiment, we prepared an Eclipse workspace packaging the classes on which the participants performed their comprehension tasks to answer the questions. The work\-space contained \textit{compilable} and functional subsets of the source code, linked to JAR files containing the rest of the compiled code of the systems. It also included a timer, the TLX program, brief tutorials on the use of Eclipse and about the systems, and a post-mortem questionnaire. No participants knew the systems on which they performed their tasks, thus we eliminated the mitigating variable relative to the participants' prior knowledge of the systems.

After the experiment, we computed the percentage of correct answers for each
question by dividing the number of correct elements found by the subject by the total number of correct elements they should have found. For example, for a question about the references to a given object, if there are ten references but the participant find only four, the percentage would be forty.

\subsection{Analysis Method}
\label{sec:analysis}

For the analyses, we use the R software environment\footnote{\url{www.r-project.org}} for statistical computing and graphics in two steps:

\paragraph{Preliminary Analysis:} We apply \textit{descriptive statistics} on the dependent variables. We also make use of \textit{boxplots}\footnote{The box-plots are drawn with the box height corresponding to the 25th and 75th percentile, with the 50th percentile (the median) marked in the box. The whiskers correspond to the 10th and 90th percentile \cite{Wohlin2012}.} to analyze the data.

\paragraph{Statistical Hypothesis Test:} We test the null hypotheses using \textit{linear mixed models (LMM)} analysis method. We use this method because it allows analyzing models with random effects (as it is the case for participants in our  experiments) and data dependency due to repeated measures (as it is our case) \citep{Vegas2016}. Moreover, thanks to \textit{LMM} analysis method, we analyze the effect of the independent variables (\textit{i.e.,} \textit{experiment}, \textit{object}, and \textit{treatment}) modelled as fixed effects on the dependent variables (\textit{i.e.,} \textit{time spent}, \textit{correctness of answers}, and \textit{effort}). The built models also include a random effect for the participants in the experiments. As it is customary with statistical hypotheses tests, we accept a probability of 5\% of committing a Type-I error (\textit{i.e.} $\alpha = 0.05$).

Considering a matrix notation, LMM can be represented as: $y = X\beta + Zu + e$. Where 
$\beta$ describe the fixed effects, the variables we want to test, that is, \emph{time}, \emph{answer}, and \emph{effort}; 
$u$ describe the random effects, in this our case the \emph{GlobalID}, which correspond to a participant's ID;
$X$ and $Z$ are the respective column values;
Finally, $e$ describe the random errors.

The requirements to apply the \textit{LMM} analysis method are: (1) residuals must follow a normal distribution and (2) the mean of the residuals must be equal to zero \citep{Vegas2016}. To check the normality of residuals, we use Shapiro-Wilk W test \citep{Shapiro1965}, where a \textit{p-value} less than $\alpha = 0.05$ indicates data not normally distributed. Without normality, we use \textit{Friedman Test} on the \textit{treatment} variable. This analysis method does not allow analyzing the effect of \textit{experiment} and \textit{object}.

%% file: sec-results.tex
% !TeX root = text-main.tex
% !TeX spellcheck = en_US
% !TeX encoding = UTF-8%

\section{Results}
\label{sec:results}

\autoref{tab:ds-blob} and \autoref{tab:ds-sc} report the descriptive statistics regarding the \emph{time} spent (in seconds) during the tasks,  correctness of the \emph{answers} (in percentage), and the \emph{effort} (in percentage) used to complete the tasks.

The tasks performed on code containing the anti-pattern Blob are, on average, \emph{33.85\% more time consuming} when compared to code without the anti-pattern. The answers for the tasks containing the anti-pattern Blob is, on average, \emph{22.07\% less accurate} when compared to task with code without the anti-pattern. Finally, the participants, on average, reported using \emph{more effort (around 10\% more)} when performing tasks involving code that contained the anti-pattern Blob.

\begin{table}[pos=ht]
\caption{Descriptive statistics showing the aggregated results for the anti-pattern \emph{Blob}.}
\label{tab:ds-blob}
\centering
\begin{tabular*}{\tblwidth}{@{}Llrrrrr@{}}
\toprule
D.Var. & Treat. & \multicolumn{1}{l}{Mean} & \multicolumn{1}{l}{Median} & \multicolumn{1}{l}{SD} & \multicolumn{1}{l}{Min} & \multicolumn{1}{l}{Max} \\ \midrule
Time & - & 155.1 & 153.1 & 60.0 & 38.2 & 319.6 \\
Time & Blob & 234.5 & 234.9 & 81.5 & 87.5 & 434.1 \\  \addlinespace
Answer & - & 60.3 & 56.1 & 18.0 & 25.0 & 95.8 \\
Answer & Blob & 38.3 & 37.3 & 18.0 & 5.5 & 71.0 \\  \addlinespace
Effort & - & 52.6 & 52.1 & 10.8 & 34.1 & 75.7 \\
Effort & Blob & 62.4 & 63.3 & 12.0 & 37.7 & 89.5 \\ \bottomrule
\end{tabular*}
\end{table}

The tasks performed on code containing the anti-pattern Spaghetti Code are, on average, \emph{39.50\% more time consuming} when compared to code without the anti-pattern. The answers for the tasks containing the anti-pattern Spaghetti Code are, on average, \emph{25.30\% less accurate} when compared to task with code without the anti-pattern. Finally, the participants reported, on average, using \emph{more effort (around 10\% more)} when performing tasks involving code that contained the anti-pattern Spaghetti Code.

\begin{table}[pos=ht]
\caption{Descriptive statistics showing the aggregated results for the anti-pattern \emph{Spaghetti Code} (SC).}
\label{tab:ds-sc}
\centering
\begin{tabular*}{\tblwidth}{@{}Llrrrrr@{}}
\toprule
D.Var. & Treat. & \multicolumn{1}{l}{Mean} & \multicolumn{1}{l}{Median} & \multicolumn{1}{l}{SD} & \multicolumn{1}{l}{Min} & \multicolumn{1}{l}{Max} \\ \midrule
Time & - & 184.5 & 183.9 & 56.5 & 86.6 & 314.5 \\
Time & SC & 257.5 & 253.8 & 74.6 & 119.9 & 457.6 \\ \addlinespace
Answer & - & 52.8 & 52.5 & 13.0 & 29.1 & 79.1 \\
Answer & SC & 27.5 & 26.5 & 17.1 & 4.1 & 66.8 \\ \addlinespace
Effort & - & 55.6 & 56.5 & 10.3 & 35.5 & 74.7 \\
Effort & SC & 65.4 & 64.2 & 11.9 & 41.5 & 89.5 \\ \bottomrule
\end{tabular*}
\end{table}

\begin{figure*}
\centering
\begin{subfigure}{.28\textwidth}
    \includegraphics[width=1\linewidth]{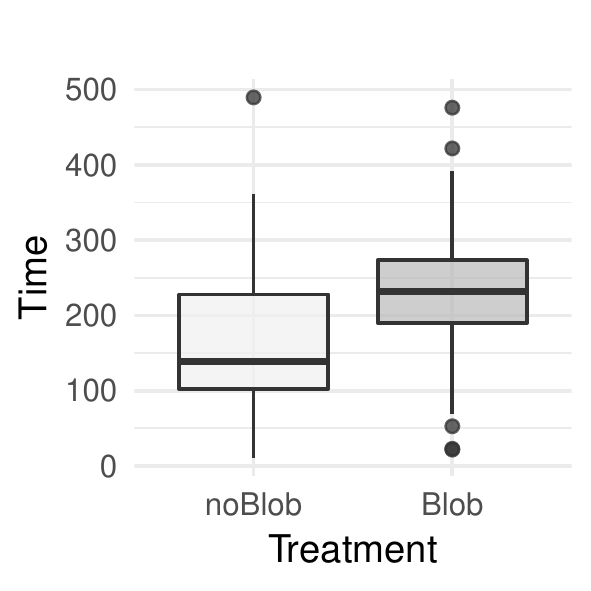}
    \caption{Blob - Time spent.}
    \label{fig:bp-time-blob}
\end{subfigure}
\begin{subfigure}{.28\textwidth}
    \includegraphics[width=1\linewidth]{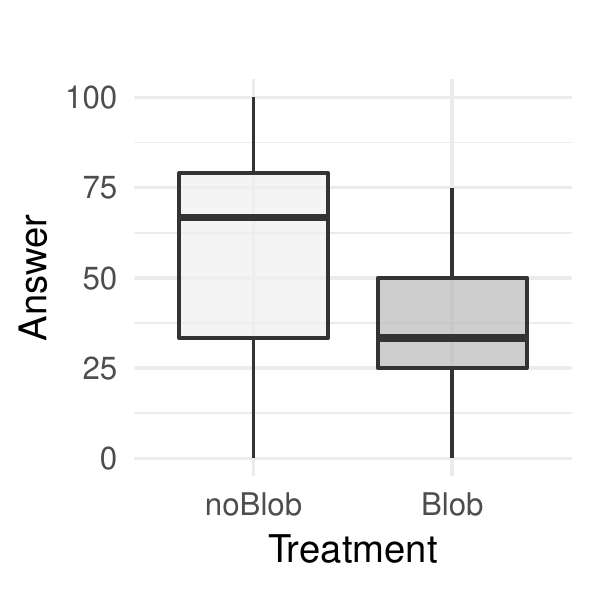}
    \caption{Blob - Correctness of Answers.}
    \label{fig:bp-answer-blob}
\end{subfigure}
\begin{subfigure}{.28\textwidth}
    \centering
    \includegraphics[width=1\linewidth]{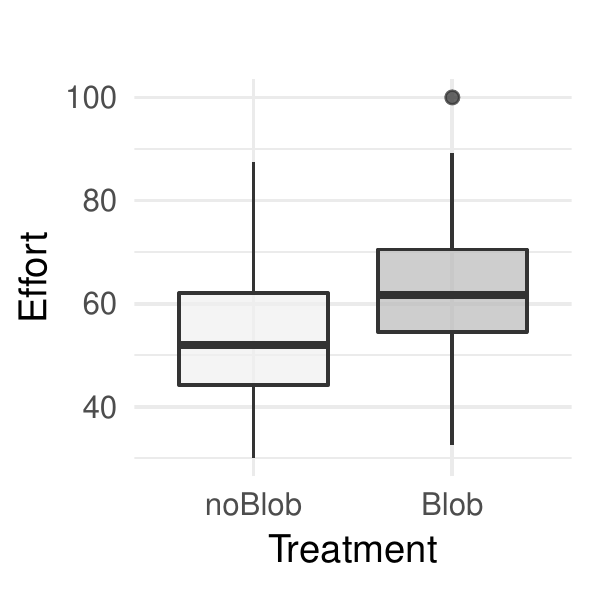}
    \caption{Blob - Overall Effort.}
    \label{fig:bp-effort-blob}
\end{subfigure}
\caption{Boxplot of the exploratory analysis with three dependent variables (time, answer, and effort) for the Blob anti-pattern.}
\label{fig:bp-blob}
\end{figure*}

\begin{figure*}
\centering
\begin{subfigure}{.28\textwidth}
    \includegraphics[width=1\linewidth]{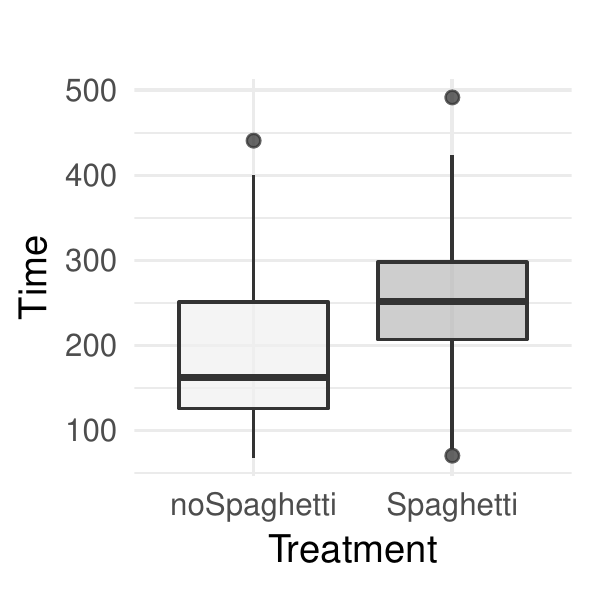}
    % \caption{Spaghetti Code - Time spent.}
    \label{fig:bp-time-sc}
\end{subfigure}
\begin{subfigure}{.28\textwidth}
    \includegraphics[width=1\linewidth]{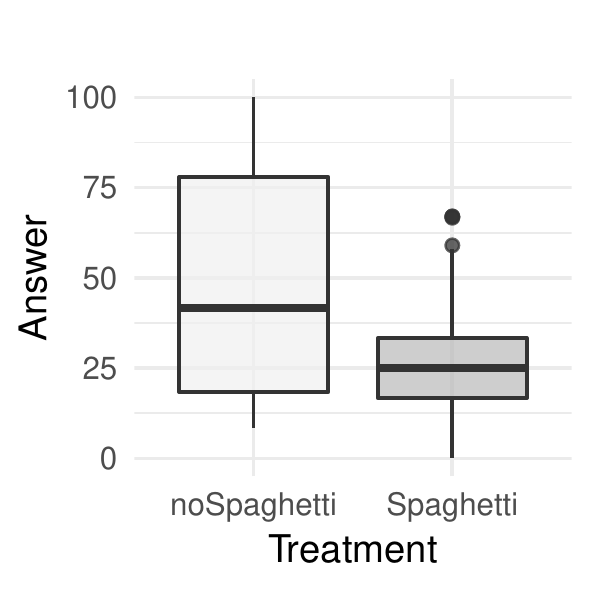}
    % \caption{Spaghetti Code - Correctness of Answers.}
    \label{fig:bp-answer-sc}
\end{subfigure}
\begin{subfigure}{.28\textwidth}
    \includegraphics[width=1\linewidth]{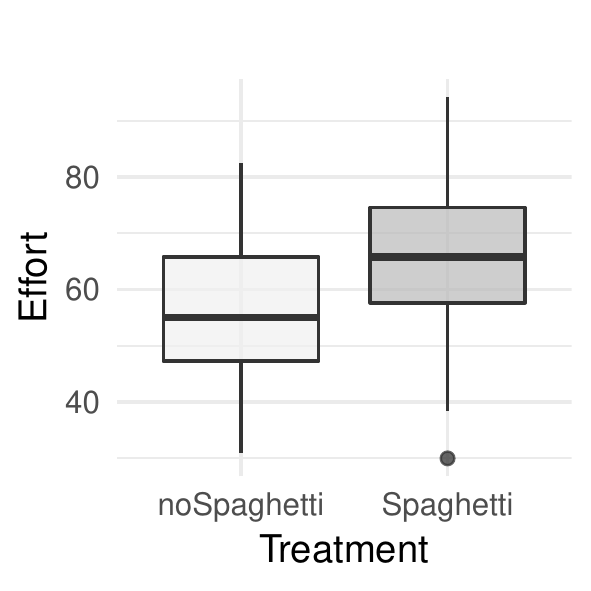}
    % \caption{Spaghetti Code - Overall Effort.}
    \label{fig:bp-effort-sc}
\end{subfigure}
\caption{Boxplot of the exploratory analysis with three dependent variables (time, answer, and effort) for the Spaghetti Code anti-pattern.}
\label{fig:bp-sc}
\end{figure*}

The \textit{boxplots} in \autoref{fig:bp-blob} and \autoref{fig:bp-sc} summarizes these results. Aside from the spread of the values regarding the variable \textit{answer} and the outliers in the variable \textit{time}, in general, the performance of the participants are better when the anti-patterns are not present.

Before applying the \textit{LMM} analysis method, we assess the normality of the residuals with the \textit{Shapiro Test}, indicating that the data are normally distributed for any dependent variable. As an example, \autoref{list:lmm} is the R script executed for the data about the Blob anti-pattern and the variable \textit{time}.

\begin{lstlisting}[float, language=R, label=list:lmm, style=code, caption=Example of the R code for the Linear Mixed Model analysis.]
# linear mixed model
LMM.Time <<- lmer(Time ~ Treatment + Experiment + Object + Treatment:Object + Treatment:Experiment + (1|GlobalID), data=dataset_blob)
# residuals check
r_blob_time <- residuals(LMM.Time)
# normality check
shap_blob_time <- shapiro.test(r_blob_time)
# perform anova in the LMM
anova.Time <- Anova(LMM.Time)
\end{lstlisting}

The variable \textit{answer}, for the experiments with the Spaghetti Code anti-pattern (\#4 and \#6) has a value less than 0.05, indicating non-normally distributed residuals (while the residuals for the other dependent variables were normally distributed). Consequently, we cannot apply the \textit{LMM} analysis method and use the \textit{Friedman Test} for the variable answer.

\autoref{tab:lmm} reports the results from the LMM analysis method (left of the vertical line) as well as those from the testing of the LMM assumptions. Each row corresponds to a LMM. For each LMM, we show the p-values for \textit{Treatment}, \textit{Experiment}, \textit{Object}, \textit{Treatment:Object} (\textit{i.e.,} the interaction between \textit{Treatment} and  \textit{Object}) and \textit{Treatment:Experiment} (\textit{i.e.,} the interaction between \textit{Treatment} and  \textit{Experiment}).

For example, the LMM built for the dependent variable \textit{time} indicates that \textit{Treatment}, \textit{Experiment}, \textit{Treatment:Object}, and \textit{Treatment:Experiment} have significant effects because their p-value is less than $0.05$. In particular, a significant effect of \textit{Treatment} means that there is a significant difference between \textit{Blob} and \textit{NoBlob}. Similarly, a significant effect of \textit{Experiment} means that there is a significant difference between the experiments, and so on.

\autoref{tab:lmm} show the results of the \textit{LMM} analysis method for Blob and Spaghetti Code anti-patterns. The variables \textit{Treatment} (anti-pattern Blob), \textit{Experiment} (Experiments \#3 and \#5), as well as their interaction (denoted as \textit{Treatment:} \textit{Experiment} in \autoref{tab:lmm}) affect the time spent, correctness of answers, and participants' effort (their p-values are less than $\alpha=0.05$). The variable \textit{Object}, \textit{e.g.,} the Java system used in the tasks, impact the correctness of the answers (while it does not impact the time spent and participants' effort). This observation could mean that some systems/tasks were more difficult to perform. Lastly, the interaction between \textit{Treatment} and \textit{Object} (\textit{e.g.,} \textit{Treatment:Object} in \autoref{tab:lmm}) affects the time spent (while the correctness of the answers and participants' effort are not affected).

\begin{table*}[width=1\textwidth,cols=4,pos=ht]
% \footnotesize
\caption{Output of the execution of the \textit{LMM} analysis method, \textit{Shapiro Test} and \textit{Mean of Residuals} for the Blob and Spaghetti Code anti-patterns. Significant effects are displayed in bold font.}
\label{tab:lmm}
\begin{tabular*}{\tblwidth}{@{}Llrrrrr|rr@{}}
\toprule
Anti-Pattern & Dep.\ Var. & Treatment & Experiment & Object & Treat.:Obj & Treat.:Exp. & Shapiro & Residual \\ \midrule
Blob & Time & \textbf{\textless~0.01} & \textbf{0.03} & 0.59 & \textbf{0.03} & \textbf{\textless~0.01} & 0.52 & 0 \\
Blob & Answer & \textbf{\textless~0.01} & \textbf{\textless~0.01} & \textbf{ \textless~0.01} & 0.22 & \textbf{\textless~0.01} & 0.82 & 0 \\
Blob & Effort & \textbf{\textless~0.01} & \textbf{0.04} & 0.74 & 0.96 & \textbf{\textless~0.01} & 0.43 & 0 \\ \addlinespace

Spaghetti & Time & \textbf{\textless~0.01} & \textbf{\textless~0.01} & 0.79 & 0.74 & \textbf{\textless~0.01} & 0.11 & 0 \\
\rowcolor[HTML]{EFEFEF}
Spaghetti & Answer & \textbf{\textless~0.01} & - & - & - & - & \textbf{ 0.02} & 0 \\
Spaghetti & Effort & \textbf{\textless~0.01} & 0.08 & \textbf{\textless~0.01} & 0.64 & \textbf{\textless~0.01} & 0.49 & 0 \\ \bottomrule
\end{tabular*}
\end{table*}

The variable \textit{Treatment} (anti-pattern Spaghetti Code) affects the time spent, correctness of answers, and participants' effort. We also observe that the variable \textit{Experiment} (Experiments \#4 and \#6) and the interaction between \textit{Treatment} and \textit{Experiment} affects the time spent (while the variable \textit{Object} and its interaction with the variable \textit{Treatment} do not affect the time spent). Finally, the variable \textit{Object} has an impact on the participants' effort as well as the interaction between the variable \textit{Treatment} and \textit{Experiment} (while \textit{Experiment} and the interaction between \textit{Treatment} and \textit{Object} do not impact the participants' effort).

\subsection{Analyzing Results by Experiments}

\begin{figure}[pos=ht]
\centering
\begin{subfigure}{.4\textwidth}
    \includegraphics[width=1\linewidth]{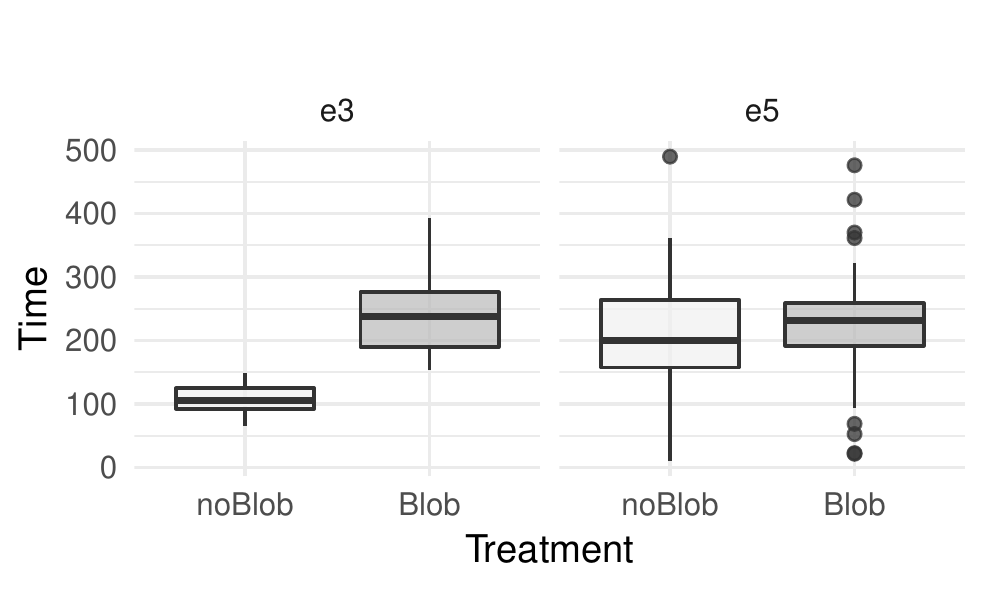}
    % \caption{Time spent.}
    \label{fig:bp-time-experiment-blob}
\end{subfigure}\hfill%
\begin{subfigure}{.4\textwidth}
    \includegraphics[width=1\linewidth]{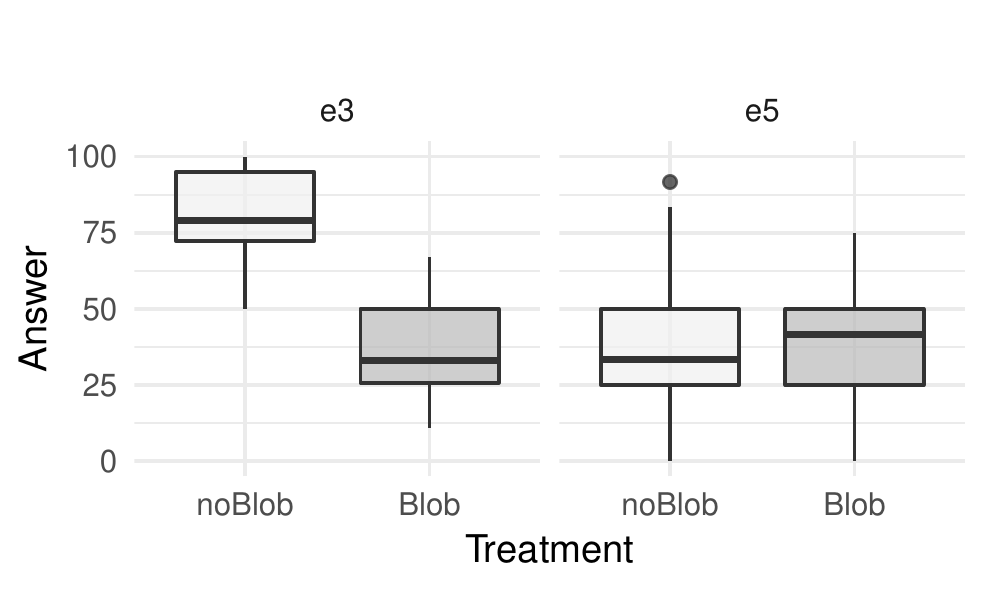}
    % \caption{Correctness of Answers.}
    \label{fig:bp-answer-experiment-blob}
\end{subfigure}\hfill%
\begin{subfigure}{.4\textwidth}
    \includegraphics[width=1\linewidth]{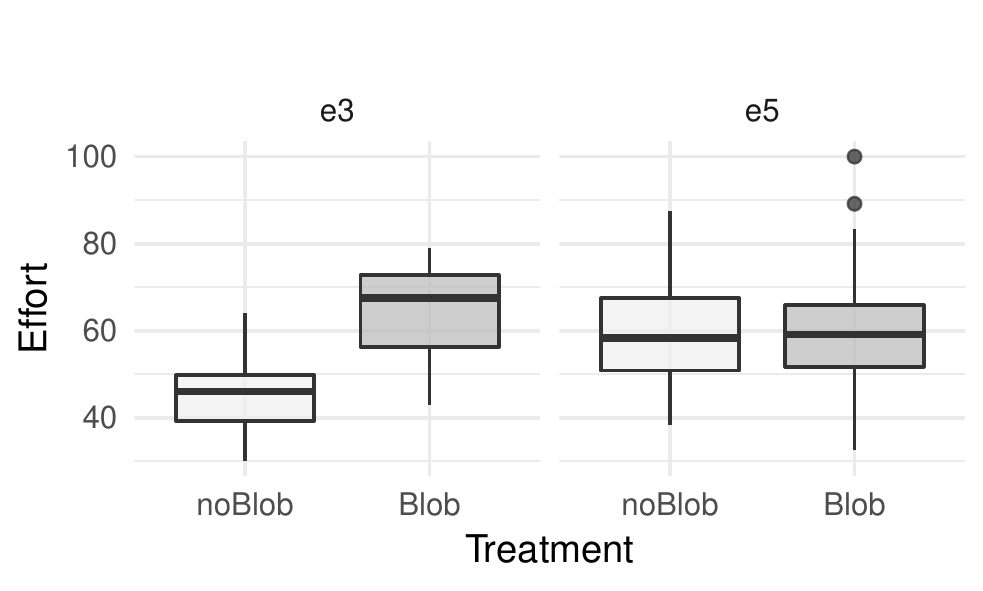}
    % \caption{Overall Effort.}
    \label{fig:bp-effort-experiment-blob}
\end{subfigure}
\caption{Boxplot of the exploratory analysis with dependent variables and the Blob anti-pattern (treatment) grouped by experiments.}
\label{fig:bp-exp-blob}
\end{figure}

\autoref{fig:bp-exp-blob} shows the boxplots grouped by experiments using the \textit{Blob} anti-pattern. Although the experiments had the same setup and design, the results present a clear gap in their values. Experiment \#3, with the tasks performed with the anti-pattern Blob, shows \emph{inferior performance} from the participants in all variables (time, accuracy, and effort) while this difference is not visible in Experiment \#5.

\begin{figure}[pos=ht]
\centering
\begin{subfigure}{.33\linewidth}
    \includegraphics[width=1\linewidth]{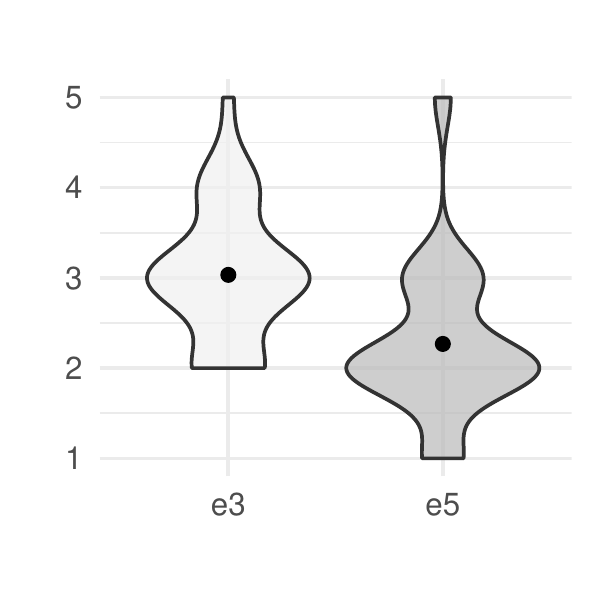}
    \caption{Soft. Eng.}
    \label{fig:vp-blob-se}
\end{subfigure}\hfill%
\begin{subfigure}{.33\linewidth}
    \includegraphics[width=1\linewidth]{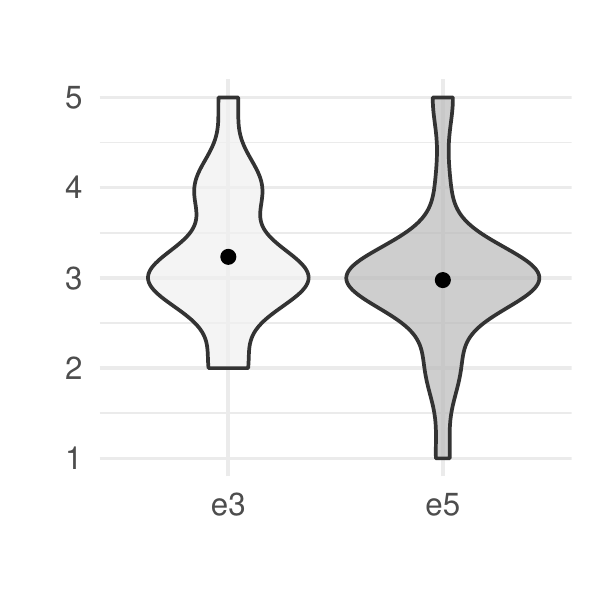}
    \caption{Java}
    \label{fig:vp-blob-java}
\end{subfigure}\hfill%
\begin{subfigure}{.33\linewidth}
    \centering
    \includegraphics[width=1\linewidth]{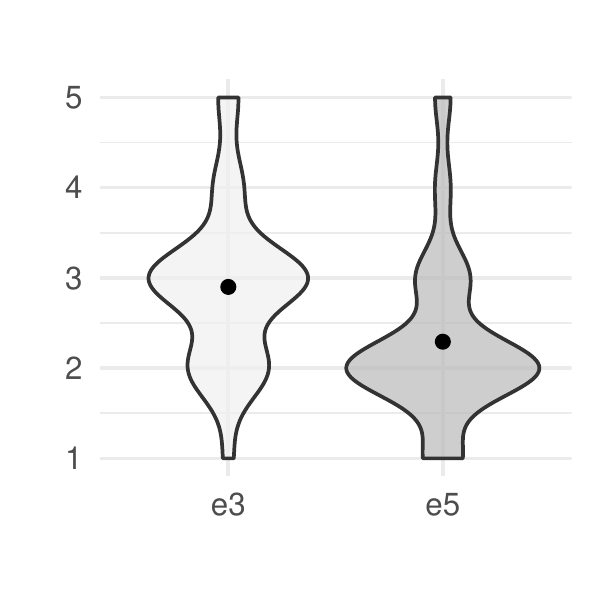}
    \caption{Eclipse}
    \label{fig:vp-blob-eclipse}
\end{subfigure}\hfill%
\caption{Violin plot showing the distribution of participants' expertise levels in \emph{Software Engineering}, \emph{Java}, and \emph{Eclipse}, for Experiments \#3 and \#5, with Blob anti-pattern.}
\label{fig:vp-blob}
\end{figure}

\begin{figure}[pos=ht]
\centering
\begin{subfigure}{.4\textwidth}
    \includegraphics[width=1\linewidth]{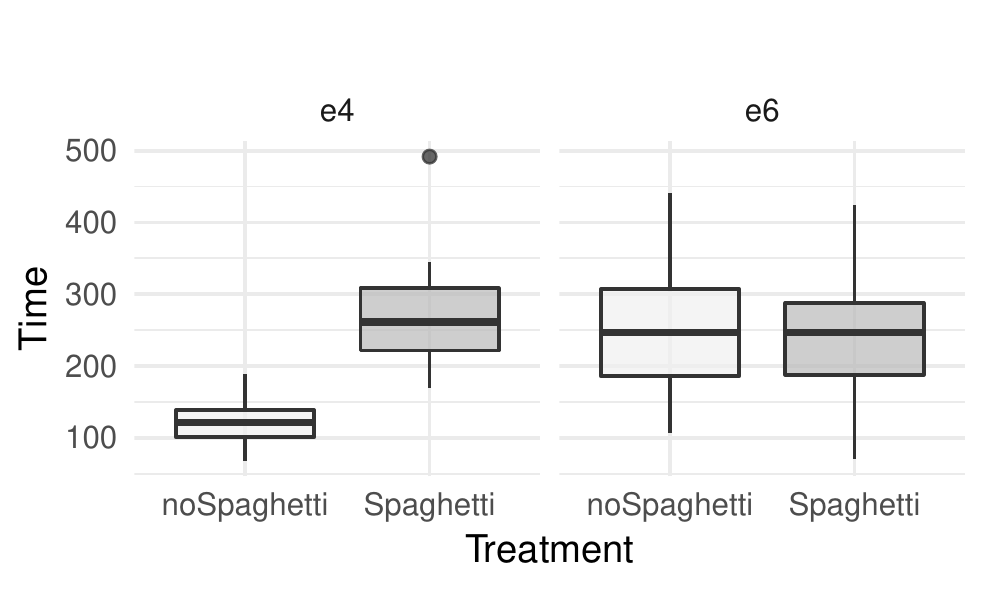}
    % \caption{Time spent.}
    \label{fig:bp-sc-time-experiment}
\end{subfigure}\hfill%
\begin{subfigure}{.4\textwidth}
    \includegraphics[width=1\linewidth]{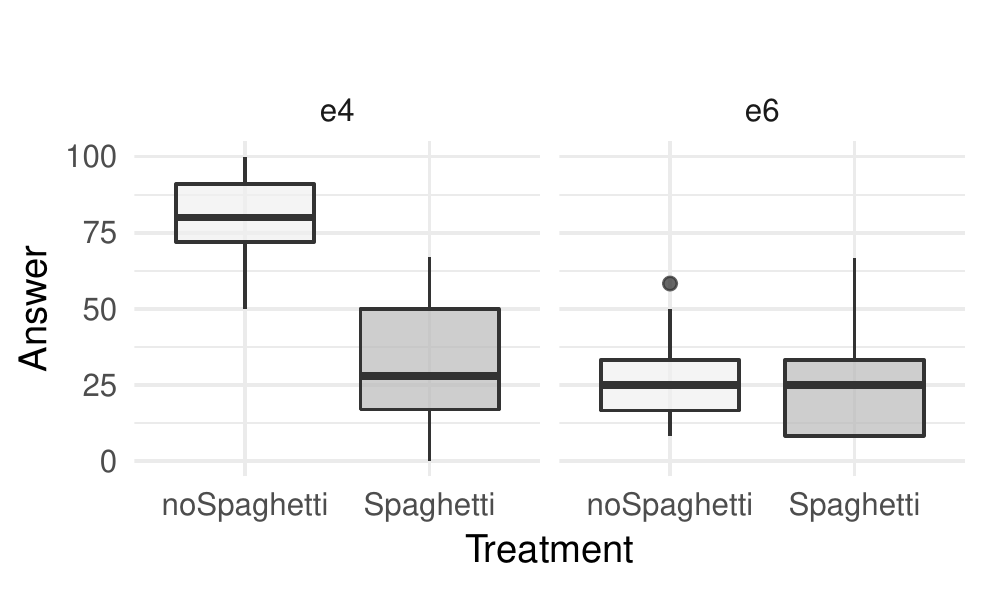}
    % \caption{Correctness of Answers.}
    \label{fig:bp-sc-answer-experiment}
\end{subfigure}\hfill%
\begin{subfigure}{.4\textwidth}
    \includegraphics[width=1\linewidth]{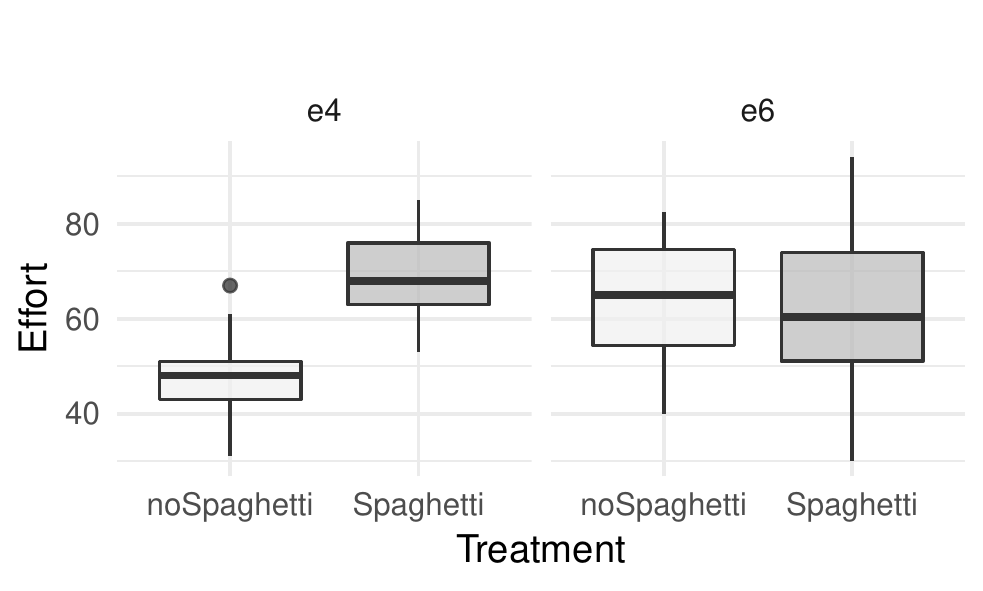}
    % \caption{Overall Effort.}
    \label{fig:bp-sc-effort-experiment}
\end{subfigure}
\caption{Boxplot of the exploratory analysis with dependent variables and the Spaghetti Code anti-pattern (treatment) grouped by experiment.}
\label{fig:bp-exp-sc}
\end{figure}

Similarly, \autoref{fig:bp-exp-sc} shows the boxplots grouped by experiments with the Spaghetti Code anti-pattern and presents disparities among values. Experiment \#4, with the Spaghetti Code, shows \emph{inferior performance} from the participants in all the variables (time, accuracy, and effort) while Experiment \#6 shows less clear difference. Also, for the variable effort, the results are slightly contradictory, with better results when the anti-pattern is present.

\begin{figure}[pos=ht]
\centering
\begin{subfigure}{.33\linewidth}
    \includegraphics[width=1\linewidth]{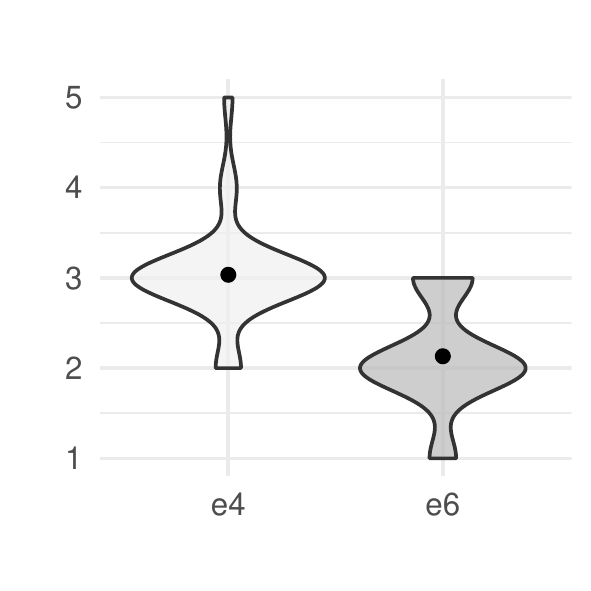}
    \caption{Soft. Eng.}
    \label{fig:vp-sc-se}
\end{subfigure}\hfill%
\begin{subfigure}{.33\linewidth}
    \includegraphics[width=1\linewidth]{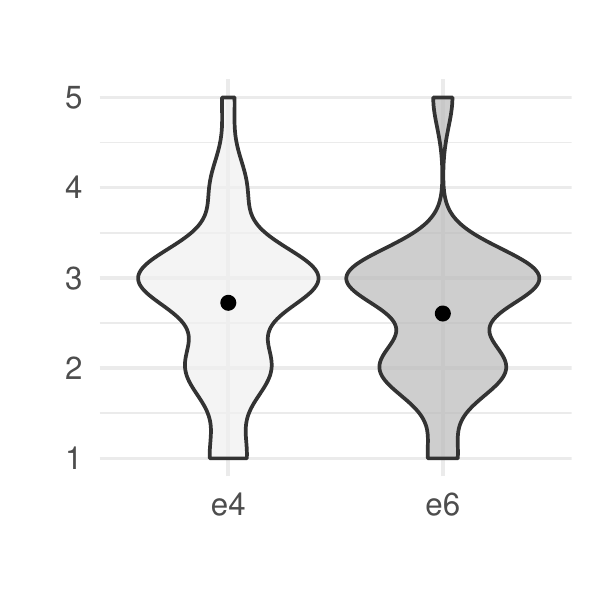}
    \caption{Java.}
    \label{fig:vp-sc-java}
\end{subfigure}\hfill%
\begin{subfigure}{.33\linewidth}
    \centering
    \includegraphics[width=1\linewidth]{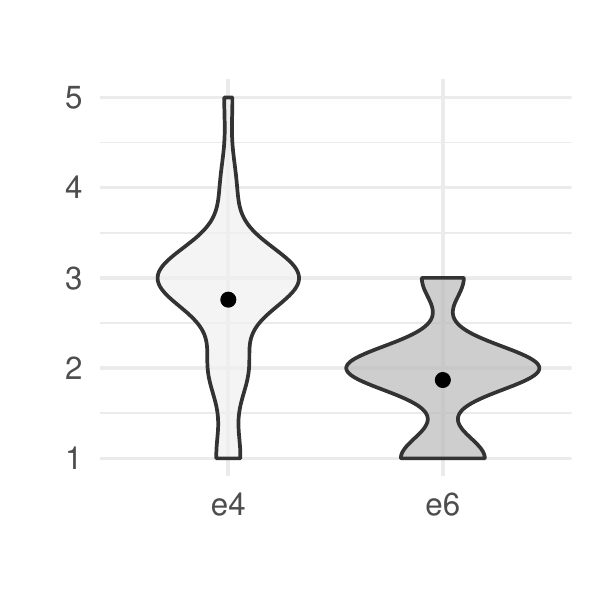}
    \caption{Eclipse.}
    \label{fig:vp-sc-eclipse}
\end{subfigure}\hfill%
\caption{Violin plots showing the distribution of participants' skills levels in \emph{Software Engineering}, \emph{Java}, and \emph{Eclipse}, for Experiments \#4 and \#6, with the Spaghetti Code anti-pattern.}
\label{fig:vp-sc}
\end{figure}

\subsection{Analyzing Participants' Skills}

To explain the disparities of the results between the experiments, \autoref{fig:vp-blob} and \autoref{fig:vp-sc} show \textit{violin plots} with the skill levels of each participant in \emph{Software Engineering}, \emph{Java}, and \emph{Eclipse}, gathered in the post-mortem questionnaire.

The participants of Experiment \#5 and \#6 have lower skills levels in Software Engineering and Eclipse but similar levels for Java. Thus, there could be a relationship between the expertise and the performance of the participants in tasks with anti-patterns. For example, participants with better technical skills had better results with source code without the anti-patterns Blob and Spaghetti Code.

To check if participants in different experiments are similar regarding their skills, we used the Mann-Whitney test. The null hypothesis was that the two samples come from the same distribution, which we would reject if the p-value is less than 0.05 (5\%). We also computed the effect sizes in both samples based on dominance matrices using Cliff's $\delta$. \autoref{list:mw} is an example of the R script to applied the tests on Experiments \#4 and \#6 (Spaghetti Code) Java programming skills.

\begin{lstlisting}[float, language=R, label=list:mw, style=code, caption=Example of the R code for the Mann-Whitney and Cliff's Delta test.]
# Mann-Whitney test
wilcox.test(x = datasetExp4$Java, y = datasetExp6$Java, paired = FALSE)
# Cliff's Delta test
cliff.delta(datasetExp4$Java, datasetExp6$Java, conf.level=.95)
\end{lstlisting}

\autoref{tab:skills} shows the results for each skill: p-value (Wil\-co\-xon test), effect (Cliff's $\delta$), and magnitude\footnotemark{} (qualitative assessment of the magnitude of effect size). There is significant difference in the participants' skills regarding \emph{Software Engineering} and \emph{Java}, for both anti-patterns.

\footnotetext{Magnitude is assessed using the thresholds:
$|\delta| < 0.147$ ``negligible'', $|\delta| < 0.33$ ``small'', $|\delta| < 0.474$ ``medium'', otherwise ``large''.}

\begin{table}[pos=ht]
\caption{Results of the Mann-Whitney and Cliff's $\delta$ tests. The values in bold font means that the null hypothesis was rejected and, therefore, that the samples are different.}
\label{tab:skills}
\begin{tabular*}{\tblwidth}{@{}lllrrl@{}}
\toprule
A. P. & Exp. & Skill & P-value & Effect & Magnitude \\ \midrule
Blob & 3 \& 4 & SE & \textbf{0.00} & 0.70 & large \\
Blob & 3 \& 4 & Java & 0.15 & 0.17 & small \\
Blob & 3 \& 4 & Eclipse & \textbf{0.00} & 0.42 & medium \\ \addlinespace
Spaghetti & 4 \& 6 & SE & \textbf{0.00} & 0.70 & large \\
Spaghetti & 4 \& 6 & Java & 0.45 & 0.10 & negligible \\
Spaghetti & 4 \& 6 & Eclipse & \textbf{0.00} & 0.59 & large \\ \bottomrule
\end{tabular*}
\end{table}

\input{sec-results-abbes}

\begin{tcolorbox}[title=Summary: Results]
The results of the \textit{descriptive statistics}, boxplots, and LMM analysis method show that either anti-pattern Blob or Spaghetti Code, when present in a java source code, decreases the developers' productivity by increasing the time spent in the tasks, reducing the accuracy of their answers, and increasing their needed effort.
\end{tcolorbox}

%% file: sec-results-abbes.tex
% !TeX root = text-main.tex
% !TeX spellcheck = en_US
% !TeX encoding = UTF-8%

\subsection{Results from \citet{Abbes2011}} \label{sec:results-abbes}

Previous experiments by \citet{Abbes2011}, which used only one instance of the anti-patterns, concluded that there is no significant impact on the source code understandability. Yet, the combination of both anti-patterns, Blob and Spaghetti Code, did decrease understandability.

\begin{figure}[pos=ht]
\centering
\begin{subfigure}{.24\textwidth}
    \includegraphics[width=1\linewidth]{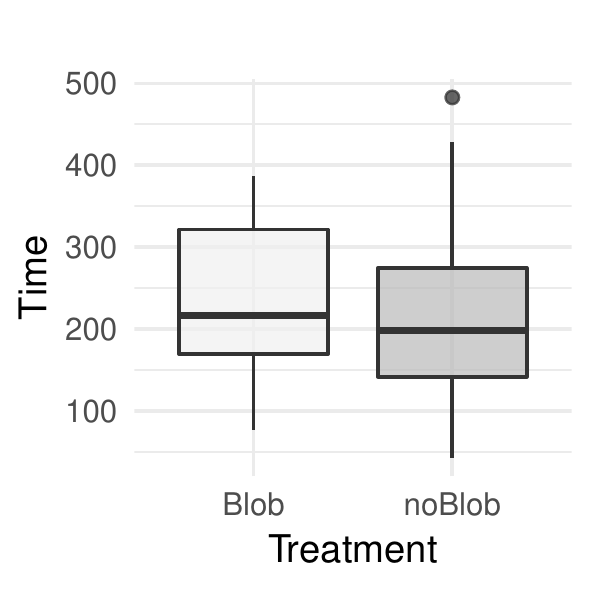}
    \caption{Time spent.}
    \label{fig:bp-time-blob-exp1}
\end{subfigure}\hfill%
\begin{subfigure}{.24\textwidth}
    \includegraphics[width=1\linewidth]{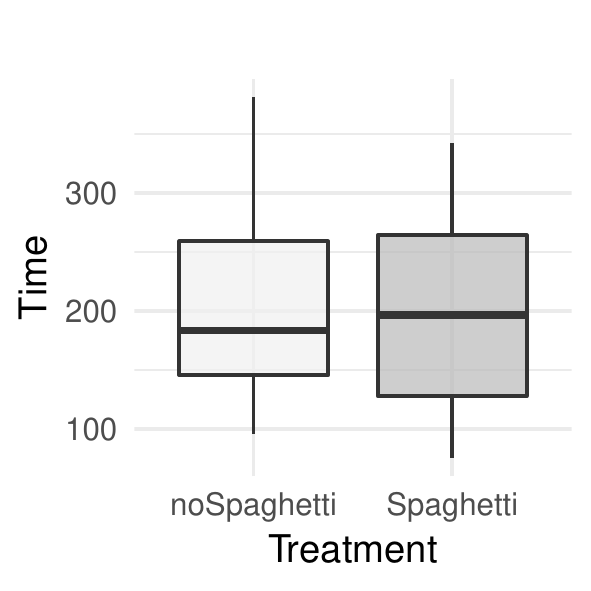}
    \caption{Time spent.}
    \label{fig:bp-time-sc-exp2}
\end{subfigure}\hfill%
\begin{subfigure}{.24\textwidth}
    \includegraphics[width=1\linewidth]{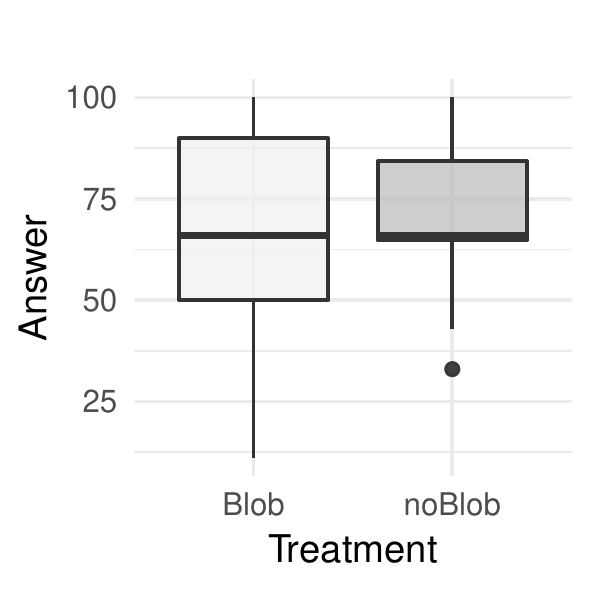}
    \caption{Correctness of Answers.}
    \label{fig:bp-answer-blob-exp1}
\end{subfigure}\hfill%
\begin{subfigure}{.24\textwidth}
    \includegraphics[width=1\linewidth]{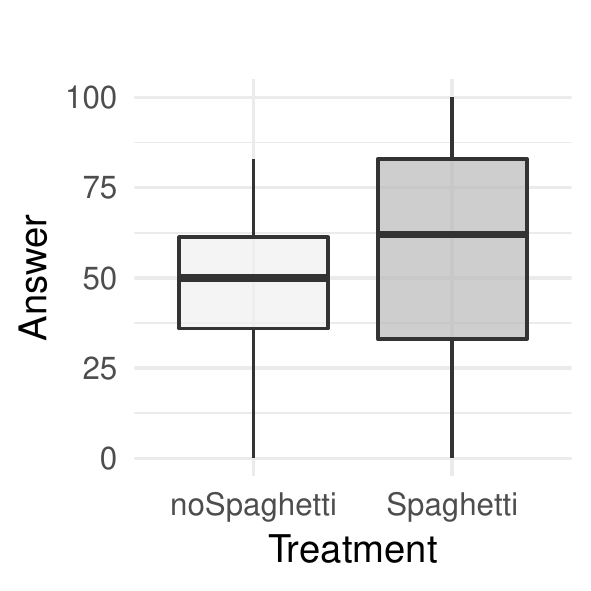}
    \caption{Correctness of Answers.}
    \label{fig:bp-answer-sc-exp2}
\end{subfigure}\hfill%
\begin{subfigure}{.24\textwidth}
    \centering
    \includegraphics[width=1\linewidth]{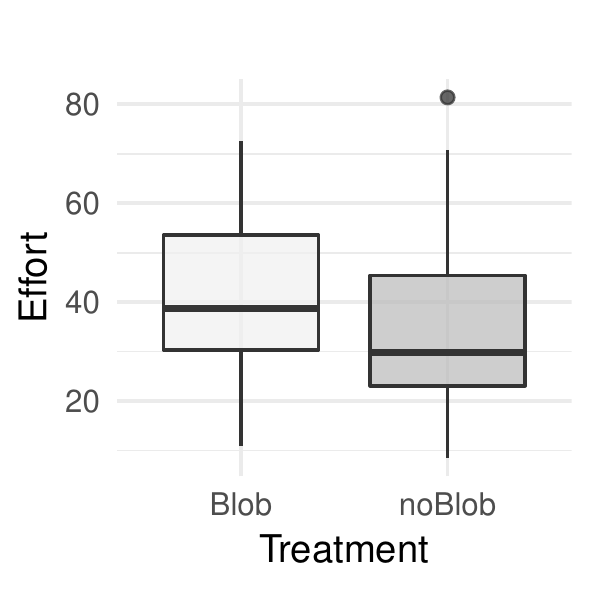}
    \caption{Overall Effort.}
    \label{fig:bp-effort-blob-exp1}
\end{subfigure}\hfill%
\begin{subfigure}{.24\textwidth}
    \centering
    \includegraphics[width=1\linewidth]{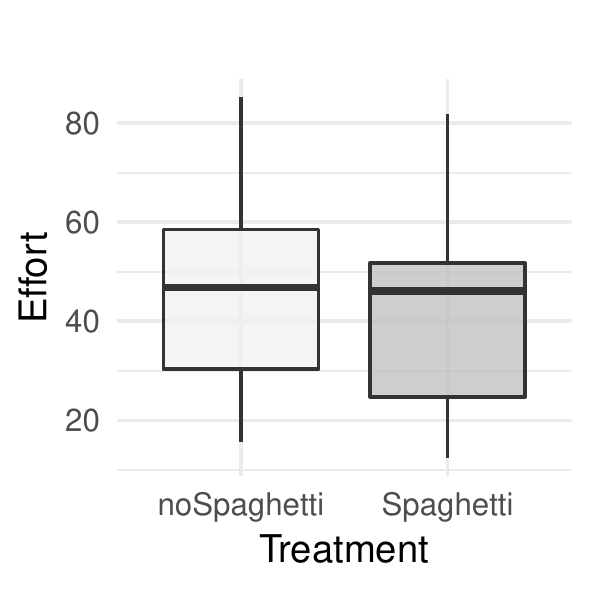}
    \caption{Overall Effort.}
    \label{fig:bp-effort-sc-exp2}
\end{subfigure}
\caption{Boxplot of the analysis with the dependent variables (time, answer, and effort) for Experiment \#1 and \#2 \citep{Abbes2011}.}
\label{fig:bp-exp1}
\end{figure}

\autoref{fig:bp-exp1} shows the boxplots of the results of Experiment \#1 and \#2 with Blob and Spaghetti Code anti-patterns. The difference regarding the mean is clear when compared with the new experiments shown in \autoref{fig:bp-blob} and \autoref{fig:bp-sc}.

Nonetheless, the results of our new experiments, with two anti-patterns instead of only one, did change the outcome. The combinations of the anti-patterns from \citep{Abbes2011}, Blob + Spaghetti Code, and the combinations of more than one occurrence of the same anti-patterns, Blob + Blob and Spaghetti Code + Spaghetti Code, harm the understandability of the source code for all the tested variables.

%% file: sec-discussion.tex
% !TeX root = text-main.tex
% !TeX spellcheck = en_US
% !TeX encoding = UTF-8%

\section{Discussion}
\label{sec:discussion}

The results from the experiments with the Blob anti-pattern show that the null hypotheses can be rejected for all dependent variables. Thus, we conclude that the Blob anti-pattern, when co-occurring twice in source code, impacts the participants' performances by increasing the time spent when performing their tasks, lowering the correctness of their answers, and increasing their effort.

Similarly, results from the experiments with the Spaghetti Code anti-pattern indicate that the null hypotheses can be rejected for all dependent variables: when co-occurring twice in source code, it also impacts the participants' performances by increasing their time spent on the tasks, lowering the correctness of their answers, and increasing their efforts.

\paragraph{Time Spent:} The time spent in any task is always a concern and usually is used as a metric of productivity, \textit{e.g.,} to estimate how long it will take to implement a new feature or fix a bug. Our results show that the time spent could be reduced by 33\% if the source code does not contain the Blob anti-pattern and by 39\% without the Spaghetti Code anti-pattern.

\paragraph{Correctness of Answers:} Developers often face source code that is hard to understand, especially when working with legacy code or third-party libraries. The poor understandability of the source code may lead to occurrences of bugs, as developers misunderstand the source code. Our results bring evidence that, when a source code contains two co-occurrences of the Blob anti-pattern or Spaghetti Code, the developers' accuracy could be reduced by 22\% and 25\%.

\paragraph{Participants' effort:} The effort used to complete a task is also a metric commonly used to measure developers' productivity. The more effort a task demands, the harder it is. Our results show that the anti-patterns Blob and Spaghetti Code could increase by 10\% developers' efforts.

\paragraph{Differences between Anti-patterns} The difference is small between the anti-patterns used in this study. However, the Spaghetti Code has a slightly bigger impact on the understandability of the source code, resulting in about 6\% more time spent and 3\% less correct answers.

\paragraph{Differences among Participants} When grouped by experiments, the results showed different outcomes. The difference in the participants' performance is clear in Experiments \#3 and \#4 (Canada) while is quite similar in the Experiments \#5 and \#6 (Italy). The former even has a contradictory better performance on effort regarding Spaghetti Code. The participants' expertise is similar for Java but different for Software Engineering and Eclipse. A first observation would be that skilled programmers work better with clean/organized code. However, we argue that we need more robust methods to model the participants' skills and measure their performance.

\paragraph{Other Consequences} The previous extended work by \citet{Abbes2011} demonstrated that a single occurrence of any anti-pattern, Blob or Spaghetti Code, does not affect source-code understanding. However, when both were mixed, understandability decreased. These latter results are confirmed by this present work, using two occurrences of each anti-pattern in the source code.

These previous observations raise the question whether the problem is the number of occurrences of anti-patterns, not necessarily their types: Is Blob + Blob more impactful than Blob + Spaghetti Code? Future work is neeed to answer this question even if the presence of the anti-patterns and their effect on the participants' performance show the importance of well-designed source code.

Understanding the impact of Blob and Spaghetti Code anti-patterns on program comprehension is important from the points of view of both researchers and practitioners. For researchers, our results bring further evidence to support the conjecture in the literature on the negative impact of anti-patterns on the quality of software systems. For practitioners, our results provide concrete evidence that they should pay attention to software systems with high numbers of classes participating in anti-patterns, because these anti-patterns reduce  understandability and, consequently, increase their systems' aging \citep{Parnas1994}. Our results also support \emph{a posteriori} the removal of anti-patterns as early as possible from systems and, therefore, the importance and usefulness of anti-patterns detection techniques and related refactorings.

\begin{tcolorbox}[title=Summary: Discussion]
The data provides preliminary evidence that anti-patterns do not affect developers with low skill level but  a source code without anti-pattern improves the developers' performance. It thus confirms the recommendation of following good practices and patterns to keep the code ``clean'' and applying refactorings to maintain it, especially for the Spaghetti Code anti-pattern.
\end{tcolorbox}

%% file: sec-threats.tex
% !TeX root = text-main.tex
% !TeX spellcheck = en_US
% !TeX encoding = UTF-8%

\section{Threats to Validity}
\label{sec:threats}

Some threats limit the validity of our study. We now discuss these threats and how we alleviate or accept them following common guidelines provided in \cite{Wohlin2012}.

\subsection{Construct Validity}

Construct validity threats concern the relation between theory and observations. In this study, they could be due to measurement errors. We use times and percentages of correct answer to measure the participants' performance. These measures are objective, even if small variations due to external factors, such as fatigue, could impact their values. We also use NASA TLX to measure the participants' effort. TLX is subjective by nature because participant were asked to rate their own efforts. Thus, it is possible that some participants did not answer truthfully.

The participants filled the questionnaire of skills by self-reporting/accessing their knowledge level, which is subjective and varied. Future research must address this issue in more details by using other techniques, for example, using eye-tracking tools.

Also, it might be possible that some occurrences of the Blob or Spaghetti Code anti-patterns are more potent in affecting source-code understandability than others. Therefore, two separate pieces of source code, both containing some anti-patterns, may have different impact of participants' performance, which should be further studied.

\subsection{Internal Validity}

We identify two threats to the internal validity of our study: learning and instrumentation.

\emph{Learning} threats should not affect our study for a specific experiment because each participant performed comprehension tasks on different systems with different questions for each system. We also took care to randomize the participants to avoid bias (\textit{e.g.,} gender bias).

\emph{Instrumentation} threats were minimized by using objective measures like times and percentages of correct answers. We observed some subjectivity in the participants' effort via TLX because, for example, one participant 100\% effort could correspond to another 50\% . However, this subjectivity illustrates the concrete participants' feeling of effort.

We strove to keep the systems/tasks as equivalent as possible, however, we accept somoe differences given the number of tasks used and the different systems.

\subsection{Conclusion Validity}

Conclusion validity threats concern the relation between the treatment and the outcome. We chose Linear Mixed Model approach to better isolate random effects, the participants in our case.
Also, we added more types of analyses in the same data, as descriptive statistics, boxplots, violin plots, and Friedman test.
We paid attention not to violate assumptions of the performed statistical tests as well not to blindly trust the LMM outcome.

\subsection{External Validity}

We performed our extended study on six different real systems belonging to different domains and with different sizes. Our design, \textit{i.e.,} providing only on average 75 classes of each system to each participant, is reasonable because, in real maintenance projects, developers perform their tasks on small parts of whole systems and probably would limit themselves as much as possible to avoid getting ``lost'' in a large code base.

We must consider that participants have different programming skills and the tasks may not suite them well. We mitigated this threat by measuring their expertise in Java and suggest that they are equivalent (see \autoref{fig:vp-blob} and \autoref{fig:vp-sc}).

%% file: sec-related-works.tex
% !TeX root = text-main.tex
% !TeX spellcheck = en_US
% !TeX encoding = UTF-8%

\section{Related Work}
\label{sec:relatedWork}

We now discuss works related to anti-patterns. \autoref{tab:rw} shows the summary of the papers in each category.

\begin{table}[width=1\linewidth,cols=4,pos=ht]
\caption{Related works grouped by research topic.}
\label{tab:rw}
\begin{tabular*}{\tblwidth}{@{}ll@{}}
    \toprule
    Anti-Pattern on... & Papers \\ \midrule
    Detection & \citep{Travassos1999, Munro2005,Oliveto2010, Marinescu2004, Moha2009, Khomh2009, Nucci2018} \\
    Quality & \citep{Olbrich2009, Chatzigeorgiou2010, Bavota2015} \\
    Fault and--or change & \citep{Khomh2009a, Khomh2012, Khomh2012, Jaafar2015, Hall2014, Palomba2017, Olbrich2010} \\
    S.L.R. & \citep{Santos2018, Sobrinho2018} \\
    Maintenance & \citep{Ignatios2004, Deligiannis2004, Yamashita2012, Sjoberg2013} \\
    Refactoring and understand. & \citep{Bois2006} \\
    Origin in projects & \citep{Tufano2017, Palomba2018} \\
    Developers reaction & \citep{Palomba2018a} \\
    Mobile energy consumption & \citep{Palomba2019} \\
    Impact in design & \citep{Oizumi2016} \\ \bottomrule
\end{tabular*}
\end{table}

Anti-patterns are poor solutions to recurrent design problems. \citet{Webster1995} wrote the first book related to anti-patterns in object-oriented development, which discussed political, conceptual, implementation, and quality-assurance problems. \citet{Fowler2012} defined 22 code smells and suggested refactorings that developers can apply to remove the code smells. \citet{Mantyla2003} and \citet{Wake2004} proposed classifications for code smells. \citet{Brown1998} described 40 anti-patterns, including the Blob and Spaghetti Code. \citet{Riel1996} defined 61 heuristics characterizing good object-oriented programming to assess the quality of systems manually and improve their designs and implementations. These books provide in-depth views on heuristics, code smells, and anti-patterns for industrial and academic audiences. In program comprehension, anti-patterns concern the design level, code smells the implementation level, and heuristics describe idioms.

% detection of code smells
Several approaches to specify and detect code smells and anti-patterns exist in the literature. They range from manual approaches, based on inspection techniques \citep{Travassos1999}, to metric-based heuristics \citep{Marinescu2004,Munro2005,Oliveto2010}, using rules and thresholds on various metrics \citep{Moha2009} or Bayesian belief networks \citep{Khomh2009}. Commercial tools such as Borland Together\footnote{\url{http://www.borland.com/us/products/together}} and AI Reviewer\footnote{\url{http://www.aireviewer.com/}} also offer the automatic detection of some code smells and anti-patterns.

% impact AP on quality
Many works investigated the impact of anti-patterns on software quality. \citet{Olbrich2009} analyzed anti-patterns in Lucene and Xerces and found that classes subjected to the Blob and Shotgun Surgery anti-patterns have a higher change frequency than other classes. Similarly, \citet{Chatzigeorgiou2010} investigated the evolution of classes with the Long Method, Feature Envy, and State Checking anti-patterns in two open-source systems and reported that a significant proportion of these anti-patterns were introduced during the addition of new methods in the systems. They also observed that these anti-patterns remain in the systems for long periods of time and that their removals are often due to adaptive maintenance rather than refactorings. \citet{Bavota2015} who analyzed 12,922 refactorings in three open-source systems also reported that refactorings are rarely aimed at removing anti-patterns.

% impact AP on change proneness
\citet{Khomh2009a} investigated the impact of several code smells on the change-proneness of classes in Azureus and Eclipse. They showed that, in general, the likelihood for classes with code smells to change is higher than classes without. \citet{Khomh2012} investigated the relation between the presence of anti-patterns and class change- and fault-proneness. They detected 13 anti-patterns in 54 releases of ArgoUML, Eclipse, Mylyn, and Rhino, and analyzed the likelihood that a class with an anti-pattern changes in the future, in particular to fix a fault. They concluded that classes participating in anti-patterns are significantly more likely to be changed and to be involved in fault-fixing changes than other classes.

\citet{Khomh2012} also investigated the kind of changes experienced by classes with anti-patterns. They considered two types of changes: structural and non-structural. Structural changes are changes that would alter a class interface while non-structural changes are changes to method bodies. They concluded that structural changes are more likely to occur in classes participating in anti-patterns. \citet{Jaafar2015} investigated the change- and fault-proneness of classes sharing a static or co-change dependency with an anti-pattern class and found that classes that are dependent on an anti-pattern class are more fault-prone than others.

% AP on faults
\citet{Hall2014} investigated the relationship between faults and five anti-patterns (Data Clumps, Switch Statements, Speculative Generality, Message Chains, and Middle Man) in three software systems (Eclipse, ArgoUML, and Apache Commons) and showed that some smells do indicate fault-prone code but the effect that these smells have on faults is small.
They also found that some smells even lead to a small reduction in fault-proneness.

% SLR on sw dev
\citet{Santos2018} carried out a systematic literature review on code smells impact in software development. They reported a strong correlation between smells and software attributes, like effort in maintenance. They observed a low human agreement on smell detection. They concluded ``that there are not strong evidences motivating the adoption of the code smells to evaluate the design quality.''.

% AP on maintenance
Previous work also investigated the impact of anti-patterns on software maintenance activities. \citet{Ignatios2004, Deligiannis2004} proposed the first quantitative study of the impact of anti-patterns on software development and maintenance activities. They performed a controlled experiment with 20 students and two systems to understand the impact of Blob classes on the understandability and maintainability of software systems. They reported that Blob classes affect the evolution of the system designs and the participants' use of inheritance. They did not assess the impact of Blob classes on the participants' understanding and their ability to perform successfully comprehension tasks on these systems.

\citet{Yamashita2012} investigated the extent to which code smells affect software maintainability and observed that code-smell definitions alone do not help developers evaluate the maintainability of a software system. They concluded on the need to combine different analysis approaches to achieve more complete and accurate evaluations of the overall maintainability of a software system.

% Refactor of AP on understandability
\citet{Bois2006} showed through a controlled experiment with 63 graduate students that the decomposition of Blob classes into a number of collaborating classes using well-known refactorings improves their understandability. They asked students to perform simple maintenance tasks on God classes and their decompositions. They found that the students had more difficulties understanding the original Blob classes than certain decompositions. However, they did not report any objective notion of ``optimal comprehensibility''.

% SLR on AP
\citet{Sobrinho2018} performed a literature review of code smell research conducted between 1990 and 2017.
% (i) Bad smell types (which); (ii) Interest on smells over time (when); (iii) Aims, findings and settings (what); (iv) Researchers (who); (v) Distribution of papers among venues (where).
The results show that
Duplicate Code is the most studied anti-pattern while the other types are not well spread, also, this research topic has been rising and attracting different researchers over the time.
There are multiple interpretations concerning anti-patterns concept and the lack of open science (standards and open data/tools) lead to contradictory results.
There is a big concentration of papers regarding Duplicate Code in some venues, which can explain the gap between the research groups.

% from reviewer 1

% Introduction of AP on projects
\citet{Tufano2017} analyzed 200 open source projects to investigate ``when and why code smells are introduced in software projects, how long they survive, and how they are removed''.
After analyzing more than 10,000 commits, they argued that smells are introduced when the artifact is created and not when it evolves. Moreover, 80\% of these smells remains in the source code, while only 9\% are removed through direct refactoring.

% AP on impact of faults and changes
\citet{Palomba2017} assessed the impact of code smells on change- and fault-proneness by empirically analyzing 395 releases of 30 Java open source projects. Among the 17,350 instances of 13 different code smell types, they found that (1) the most diffused smells are related to size, like Long Method, Spaghetti Code; (2) classes affected by code smells are more suitable to changes and faults; (3) removing code smells does affect change-proneness in the class but it is beneficial to the occurrence of changes.

% origins of AP in projects
\citet{Palomba2018} investigated the nature of code smell co-occurrences in 395 projects. The results showed that 59\% of the smelly classes are affected by more than one code smell. Among them, there are six code smell types frequently co-occurring together, being the \textit{Message Chain} the most common.
Still, the method-level smells may be the root cause of the class-level ones.
Finally, the co-occurring smells tend to be removed together.

% developers reaction on AP
\citet{Palomba2018a} conducted a quantitative and qualitative study to check how developers react to smells depending of the tool used to its detection, that is, textual-base or structural-base. Among the results, they reported that both types are considered equally harmful for the project.

% AP of mobile energy consumption
\citet{Palomba2019} conducted an experiment with 9 Android-specific code smells on 60 Android apps to assess its energy consumption. The results showed the existence of four specific energy code smells which, when refactored, increase the energy efficiency.

% on detecting APs
\citet{Nucci2018} replicate a study using 32 machine-learning techniques to detect four types of code smell by extending the dataset. Results showed differences in the previous output, which has 95\% accuracy, being 90\% less accurate.

% reviewer 3

% AP on change and fault proneness
\citet{Olbrich2010} analyzed historical data from 7 to 10 years of the development of three open-source software systems. They found that after controlling for differences in class size, Brain and God Classes, are less change- and fault-prone than other classes.

% AP on maintenance
\citet{Sjoberg2013} hired six Java developers which modified 298 Java files in different tasks in the four systems while measuring the time spent. They essentially found that code smells do not have a consistent effect on maintenance effort if differences in file size are controlled for.

% AP on impact on design
\citet{Oizumi2016} analyzed more than 2,200 agglomerations (inter-related code anomalies) found in seven software systems of different sizes and from different domains. Their findings suggest that a single code smell seldom indicates a design problem, but if two or more smells co-occur, they indicate design problems with high accuracy.

\begin{tcolorbox}[title=Summary: Related Works]
\autoref{tab:rw} shows that the literature on anti-patterns has its roots in detection and fault/change proneness. Few studies investigated the impact of anti-patterns on the understandability. In this paper, we built on previous work and propose experiments assessing the impact of the Blob and Spaghetti Code on source-code understandability.
\end{tcolorbox}

%% file: sec-conclusion.tex
% !TeX root = text-main.tex
% !TeX spellcheck = en_US
% !TeX encoding = UTF-8%
\section{Conclusion and Future Work}
\label{sec:conclusion}

Anti-patterns were conjectured in the literature to negatively impact the quality of systems. Some studies empirically investigated the impact of anti-patterns on program comprehension. We revisited the studies on the impact of the Blob and Spaghetti Code on program comprehension by complementing a previous work by \citet{Abbes2011}.

We designed and conducted two replications with 133 participants and 372 program-compre\-hen\-sion tasks to assess the impact of combinations of occurrences of the Blob and Spaghetti Code anti-patterns. We measured developers' performances using: (1) the time that they spent performing their tasks; (2) their percentages of correct answers; and, (3) the NASA task load index for their effort.

We showed, in complement to the results of the previous work, that either two instances of the Blob anti-pattern or two instances Spaghetti Code anti-pattern negatively impact the understandability of source code by increasing the time spent by participants in performing the tasks, by reducing the correctness of their answers, and by increasing their efforts.

Consequently, developers should be wary with growing numbers of Blob and Spaghetti Code anti-patterns in their systems and, therefore, refactor these occurrences out of the source code of their systems. Researchers should study \emph{combinations} of anti-patterns (and other practices, like code smells) rather than single anti-patterns, one at a time.

Future work includes investigating the relation between the number of occurrences of anti-patterns and  understandability. We plan to investigate how to quantify the impact of occurrences of an anti-pattern on understandability. We also plan to replicate these studies in other contexts, with other participants, other anti-patterns, and other systems.